\documentclass[longauth]{aa}
\usepackage{graphicx}
\usepackage{txfonts}
\usepackage[switch]{lineno}
\usepackage{booktabs}
\usepackage[table]{xcolor}
\usepackage{colortbl}
\usepackage{orcidlink}
\setcitestyle{aa}

\renewcommand{\emph}[1]{\protect\textit{#1}}
\usepackage{cancel}
\usepackage{ulem}
\usepackage{color}
\usepackage{natbib, twoopt}
\bibpunct{(}{)}{;}{a}{}{,}    
\definecolor{cobalt}{rgb}{0.06, 0.2, 0.65}
\hypersetup{
  colorlinks,
  citecolor=cobalt,
  linkcolor=cobalt,
  urlcolor=cobalt,
}
\makeatletter
  \newcommandtwoopt{\citeads}[3][][]{\href{http://adsabs.harvard.edu/abs/#3}%
    {\def\hyper@linkstart##1##2{}%
     \let\hyper@linkend\@empty\citealp[#1][#2]{#3}}}
  \newcommandtwoopt{\citepads}[3][][]{\href{http://adsabs.harvard.edu/abs/#3}%
    {\def\hyper@linkstart##1##2{}%
     \let\hyper@linkend\@empty\citep[#1][#2]{#3}}}
  \newcommandtwoopt{\citetads}[3][][]{\href{http://adsabs.harvard.edu/abs/#3}%
    {\def\hyper@linkstart##1##2{}%
     \let\hyper@linkend\@empty\citet[#1][#2]{#3}}}
  \newcommandtwoopt{\citeyearads}[3][][]%
    {\href{http://adsabs.harvard.edu/abs/#3}
    {\def\hyper@linkstart##1##2{}%
     \let\hyper@linkend\@empty\citeyear[#1][#2]{#3}}}
\makeatother

\renewcommand{\emph}[1]{\protect\textit{#1}}
\usepackage{adjustbox}

\begin{document} 

   \title{The Rosetta Stone project}
   
   \subtitle{III. ALMA synthetic observations of fragmentation in high-mass star-forming clumps}

   \authorrunning{A. Nucara et al.}
   \titlerunning{Rosetta Stone. Paper III}

   \author{
          Alice~Nucara\inst{\ref{iaps}}\fnmsep\inst{\ref{ToV}}\fnmsep\thanks{alice.nucara@inaf.it}\orcidlink{0009-0005-9192-5491}
          \and
          Alessio~Traficante\inst{\ref{iaps}}\orcidlink{0000-0003-1665-6402}
          \and 
          Ugo~Lebreuilly\inst{\ref{Saclay}}\orcidlink{0000-0001-8060-1890}
          \and
          Ngo-Duy~Tung\inst{\ref{Saclay}}\orcidlink{0009-0009-8545-2682}
          \and
          Sergio~Molinari\inst{\ref{iaps}}\orcidlink{0000-0002-9826-7525}
          \and
          Patrick~Hennebelle\inst{\ref{Saclay}}\orcidlink{0000-0002-0472-7202}
          \and
          Leonardo~Testi\inst{\ref{DIFA}}\fnmsep\inst{\ref{Arcetri}}\orcidlink{0000-0003-1859-3070}
          \and Ralf~S.~Klessen\inst{\ref{Heidelberg1}}\fnmsep\inst{\ref{Heidelberg2}}\fnmsep\inst{\ref{Cfa}}\fnmsep\inst{\ref{Radcliffe}}\orcidlink{0000-0002-0560-3172}
          \and
          Veli-Matti~Pelkonen\inst{\ref{iaps}}\orcidlink{0000-0002-8898-1047}
          \and
          Adam~Avison\inst{\ref{SKA}}\orcidlink{0000-0002-2562-8609}
          \and 
          Milena~Benedettini\inst{\ref{iaps}}\orcidlink{0000-0002-3597-7263}
          \and
          Alessandro~Coletta\inst{\ref{iaps}}\orcidlink{0000-0001-8239-8304}
          \and
          Fabrizio~De~Angelis\inst{\ref{iaps}}\orcidlink{0009-0002-6765-7413}
          \and
          Davide~Elia\inst{\ref{iaps}}\orcidlink{0000-0002-9120-5890}
          \and
          Gary~A.~Fuller\inst{\ref{JBCA}}\fnmsep\inst{\ref{Koln}}\orcidlink{0000-0001-8509-1818}
          \and 
          Bethany~M.~Jones\inst{\ref{Koln}}\orcidlink{0000-0002-0675-0078}
          \and
          Seyma Mercimek\inst{\ref{JBCA}}\orcidlink{0000-0002-0742-7934}
          \and
          Chiara~Mininni\inst{\ref{iaps}}\orcidlink{0000-0002-2974-4703}
          \and
          Stefania~Pezzuto\inst{\ref{iaps}}\orcidlink{0000-0001-7852-1971}
          \and
          Thushara~Pillai\inst{\ref{Haystack}}\fnmsep\inst{\ref{Boston}}\orcidlink{0000-0003-2133-4862}\and
          Veronica~Roccatagliata\inst{\ref{DIFA}}\fnmsep\inst{\ref{Arcetri}}\orcidlink{0000-0002-4650-594}
          \and 
          Eugenio~Schisano\inst{\ref{iaps}}\orcidlink{0000-0003-1560-3958}
          \and
          Juan~D.~Soler\inst{\ref{iaps}}\orcidlink{0000-0002-0294-4465}
          \and
          Paolo~Suin\inst{\ref{Marseille}}\orcidlink{0000-0001-7044-3809}
          \and
          Claudia~Toci\inst{\ref{Arcetri}}\fnmsep\inst{\ref{ESO}}\orcidlink{0000-0002-6958-4986}
          \and
          Daniel~Walker\inst{\ref{JBCA}}\orcidlink{0000-0001-7330-8856}
          }

   \institute{INAF - Istituto di Astrofisica e Planetologia Spaziali, Via Fosso del Cavaliere 100, I-00133 Roma, Italy \label{iaps}
         \and
         Dipartimento di Fisica, Università di Roma Tor Vergata, Via della Ricerca Scientifica 1, I-00133 Roma, Italy \label{ToV}
         \and
         Université Paris-Saclay, Université Paris Cité, CEA, CNRS, AIM, F-91191, Gif-sur-Yvette, France \label{Saclay}
         \and 
         Alma Mater Studiorum Università di Bologna, Dipartimento di Fisica e Astronomia (DIFA), Via Gobetti 93/2, I-40129, Bologna, Italy \label{DIFA}
         \and
         INAF-Osservatorio Astrofisico di Arcetri, Largo E. Fermi 5, I-50125, Firenze, Italy \label{Arcetri}
         \and
         Universität Heidelberg, Zentrum für Astronomie, Institut für Theoretische Astrophysik, Albert-Ueberle-Str. 2, 69120 Heidelberg, Germany \label{Heidelberg1}
         \and 
         Universität Heidelberg, Interdisziplinäres Zentrum für Wissenschaftliches Rechnen, Im Neuenheimer Feld 205, 69120 Heidelberg, Germany \label{Heidelberg2}
         \and
         Harvard-Smithsonian Center for Astrophysics, 60 Garden Street, Cambridge, MA 02138, U.S.A. \label{Cfa}
         \and
         Radcliffe Institute for Advanced Studies at Harvard University, 10 Garden Street, Cambridge, MA 02138, U.S.A. \label{Radcliffe}
         \and
         SKA Observatory, Jodrell Bank, Lower Withington, Macclesfield, SK11 9FT, UK \label{SKA}
         \and 
         Jodrell Bank Centre for Astrophysics, Department of Physics \& Astronomy, The University of Manchester, Oxford Road, Manchester M13 9PL, UK \label{JBCA}
         \and
         Physikalisches Institut der Universität zu Köln, Zülpicher Str. 77, D-50937 Köln, Germany \label{Koln}
         \and
         Haystack Observatory, Massachusetts Institute of Technology, 99 Millstone Rd., Westford, MA 01886, USA \label{Haystack}
         \and
         Institute for Astrophysical Research, Boston University, 725 Commonwealth Avenue, Boston, MA 02215, USA \label{Boston}
         \and
         Aix Marseille Univ, CNRS, CNES, LAM Marseille, France \label{Marseille}
         \and
         European Southern Observatory (ESO), Karl-Schwarzschild-Strasse 2, 85748, Garching bei Munchen, Germany \label{ESO}
        }

   \date{Received March 26, 2025; accepted June 2, 2025}

  \abstract
   {The physical mechanisms that regulate the collapse of high-mass parsec-scale clumps and allow them to form clusters of new stars, including high-mass stars, represent a crucial aspect of star formation.}
   {To investigate these mechanisms, we developed the Rosetta Stone project: an end-to-end (simulations $\Leftrightarrow$ observations) framework that is based on the systematic production of realistic synthetic observations of clump fragmentation and their subsequent comparison with real data.}
   {In this work, we compare ALMA 1.3 mm continuum dust emission observations from the Star formation in QUiescent And Luminous Objects (SQUALO) survey with a new set of $24$ radiative magnetohydrodynamical (RMHD) simulations of high-mass clump fragmentation, post-processed using the CASA software to mimic the observing strategy of SQUALO (combining ACA and \texttt{12\,m} array).
   The simulations were initialized combining typical values of clump mass (500 and 1000 M$_\odot$) and radius ($\sim0.4$ pc) 
   with two levels of turbulence (Mach number, $\mathcal{M}$, of $7$ and $10$) and three levels of magnetization (normalized mass-to-magnetic-flux ratio, $\mu$, of $\sim3$, $10$, and $100$).
   Following the clump evolution over time with two initial random seeds projected along three orthogonal directions, we produced a collection of $732$ synthetic fields.
   On each field, we performed source extraction and photometry using the \textit{Hyper} software, as in the SQUALO project, to quantitatively characterize how the initial conditions of the clump and the environment affect the observed fragmentation properties.}
   {The synthetic observations of clump fragmentation at $\sim7000$ AU resolution revealed between $2$ and $14$ fragments per field, 
   indicating a complex fragmentation process.
   Among the initial conditions of the simulations, magnetic fields have the largest impact on the fragment multiplicity at these scales. 
   In advanced stages of clump evolution, a lower number of fragments is preferentially associated with magnetized clumps.
   The clump magnetization might also affect the clustering of fragments, favoring more tightly bound distributions when the magnetic field is stronger. 
   Fragments identified at $\sim7000$ AU correspond to individual or multiple sink particles in $\sim 75\%$ of the cases.
   This result suggests that not all identified fragments are actively forming stars.
   Both sink particles and fragments accrete mass throughout the whole clump evolution.
   This evidence favors a scenario in which fragments are not isolated from the environment and is thus consistent with results from the SQUALO survey.}
   {Our study demonstrates the importance of synthetic observations in interpreting results from interferometric observations.
   }


   \maketitle

\section{Introduction}
    Star formation, particularly in high-mass star-forming regions, is a multi-scale process (e.g., \citealt{2019MNRAS.490.3061V},  
    \citealt{2023MNRAS.522.3719L}, 
    \citealt{2023ApJ...953...40Y}). 
    It develops within giant molecular clouds and filaments, where the assembly of parsec-scale clumps takes place (e.g., \citealt{2010A&A...518L.100M}), and then progresses through a hierarchy of nested structures on scales as small as $\sim1000$ AU (e.g.,  
    \citealt{2019ApJ...886..102S},  
    \citealt{2019ApJ...886...36S}, 
    \citealt{2022A&A...662A...8M},
    \citealt{2023MNRAS.520.2306T}, 
    \citealt{2024ApJ...966..171M},
    \citealt{2024RAA....24b5009L}, 
    \citealt{2024ApJS..270....9X}, 
    \citealt{2025A&A...696A.151C}).\\
    \indent A crucial aspect of this process is represented by the physical mechanisms that regulate the collapse of clumps of mass of the order of $ \sim 10^3\, \text{M}_\odot$, ultimately leading to the formation of clusters of new stars, including future high-mass stars (e.g., 
    \citealt{2003ARA&A..41...57L}, 
    \citealt{2010A&A...518L..95H},
    \citealt{2018A&A...617A.100B},
    \citealt{2022A&A...664A..26P}). 
    Observational studies and numerical simulations have been employed to quantify the impact of gravity, turbulence, magnetic fields, and feedback mechanisms - as well as that of the potential interplay of these forces - on clump fragmentation properties.
    However, results from these works have not been conclusive, encompassing a range of viable star-formation models beyond the long-established dichotomy between the core-fed (e.g., \citealt{2003ApJ...585..850M}, \citealt{2014prpl.conf..149T}) and the clump-fed scenarios (e.g., \citealt{Bonnell1997,Bonnell2001}, \citealt{Klessen2000}, \citealt{2006MNRAS.370..488B}, \citealt{Girichidis2011,Girichidis2012},   
    \citealt{2012A&A...543L...3H},  
    \citealt{2019MNRAS.490.3061V}, 
    \citealt{2020MNRAS.496.3482P}, \citealt{2024arXiv240810406V}).\\
    \indent The most recent studies involving high-mass star-forming regions
    exhibit shared evidence favoring a scenario in which fragments are not isolated from the environment. 
    While some works report thermal Jeans fragmentation as the leading star-formation mechanism (e.g., 
    \citealt{2015MNRAS.453.3785P}, 
    \citealt{2019ApJ...886..102S},  
    \citealt{2021A&A...649A.113B}, \citealt{2024ApJ...966..171M}, \textcolor{cobalt}{Schisano et al. (submitted)}),
    others indicate that additional support from turbulence, magnetic pressure, and feedback mechanisms or a continuous mass accretion onto the fragments must be in place during clump collapse 
    (e.g., \citealt{2007prpl.conf...63B}, 
    \citealt{2009ApJ...696..268Z},  
    \citealt{2011A&A...528A..72H}, 
    \citealt{2011A&A...530A.118P}, 
    \citealt{2011ApJ...735...64W},  
    \citealt{2018A&A...615A..94F}, 
    \citealt{2023MNRAS.520.2306T}, 
    \citealt{2023MNRAS.520.3259X}, 
    \citealt{2024MNRAS.530.3445V}).
    Magnetic regulation, in particular, seems to be crucial in the case of massive clumps that exhibit a few fragments and/or the presence of super-Jeans fragments  (e.g.,
    \citealt{2015ApJ...804..141Z}, 
    \citealt{2016A&A...591A..19P},
    \citealt{2022A&A...668A.147H},
    \citealt{2023ASPC..534..193P}); 
    yet some works have indicated only partial agreement or conflicting evidences (e.g., \citealt{2021ApJ...912..159P}, \citealt{2024A&A...682A..81B}).\\
    \indent Although conflicting pieces of evidence are often due to the diversity of the explored environments, they may also be due to the spatial scales probed in the various works (e.g., \citealt{2021A&A...653A.157L}).
    Indeed, the relative importance of gravitational energy, magnetic field strength, turbulence, and radiative pressure might vary across scales, leading to scale-dependent fragmentation patterns regulated by the dominant force at the scale under analysis, and across environments (e.g., \citealt{2018A&A...615A..94F}, \citealt{2019ApJ...878...10T}, \citealt{2022A&A...662A...8M},  \citealt{2024ApJ...962..136W}).
    While defining a universal model for clump fragmentation can be particularly challenging under these circumstances, the comparison between observations and simulations
    has proven to be a useful tool for constraining star-formation mechanisms, especially when involving synthetic observations
    (e.g., 
    \citealt{2007ApJ...656..959K},
    \citealt{2011IAUS..270..511G}, 
    \citealt{2014ApJ...784...61O}, 
    \citealt{2017ApJS..233....1K},
    \citealt{2019ApJ...886...36S}, 
    \citealt{2023ApJ...950...88J}).\\
    \indent With a specific focus on viable scenarios of clump fragmentation, \cite{2020ApJ...900...82P} used synthetic observations to test the core collapse and the competitive accretion models against {\textit{Herschel}} and Atacama Large Millimeter/submillimeter Array (ALMA) observations.
    They proposed the inertial-inflow model to describe turbulent fragmentation and its potential connection to the origin of massive stars.
    \cite{2018A&A...615A..94F} carried out analogous studies to analyze the impact of the interplay of turbulence and magnetic fields within simulated clumps.
    Comparing their results with the fragmentation properties of massive clumps observed with ALMA, they suggest that higher levels of turbulence produce more fragments and that higher magnetization favors the formation of filamentary structures.
    At the same time, 
    works by \cite{2011A&A...530A.101M}, \cite{2017MNRAS.471.4111H}, and \cite{2023MNRAS.522.3548P}, for example, have proven the utility of synthetic observations in evaluating technical aspects concerning real observations.
    They used synthetic data to quantify the reliability of clump, core, and protostar mass estimates derived from submillimeter and infrared dust emission.  
    \cite{2014ApJ...783...60M}, \cite{2018A&A...617A.100B}, and \cite{2024A&A...688A..30B} used synthetic observations to further address the difference in sensitivity to the more extended emission related to single-dish and interferometeric observations, including the ability of detecting the smallest structures and the non-negligible impact of large-scale interferometric filtering. \\
    \indent Despite the advent of synthetic observations have led to significant advancements in how the star-formation process is understood (e.g., 
    \citealt{2018NewAR..82....1H},
    \citealt{2020SSRv..216...62R}),     
    the majority of works involving synthetic data are still subject to some limitations.
    In particular, the theoretical setups are typically tuned to mimic a specific environment and/or target source, hindering the chance to perform a broad and self-consistent exploration of the parameter space and therefore failing to capture the variety of real star-forming systems (e.g., \citealt{2007ARA&A..45..565M}, \citealt{2012A&ARv..20...55H}, \citealt{2016SAAS...43...85K}). 
    The comparison between observations and synthetic data itself represents an issue.
    Observational features caused by the resolution limit and by spatial filtering, as well as by the presence of background and interferometric noise, are often not well reproduced in the synthetic observations, precluding a one-to-one comparison with available real data. 
    To effectively address the physics governing star-forming sites at the scales probed by the observations, it is therefore necessary to apply a rigorous post-processing routine to the simulations and ensure that these also mimic the observational features (instrumental effects and observational biases) of the dataset they will be compared against. 
    \begin{figure*}
        \centering
        \includegraphics[width=\textwidth]{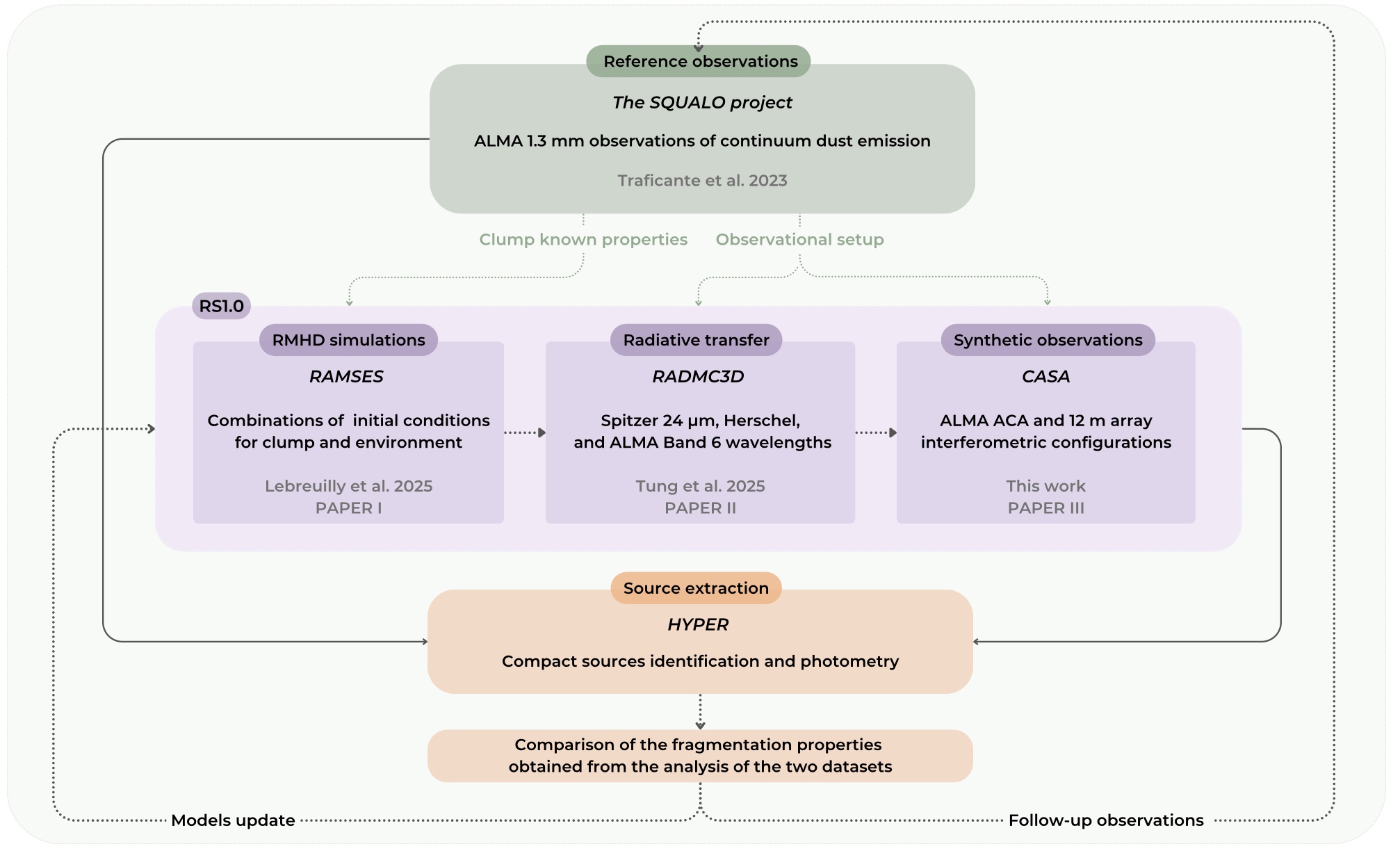}

        \caption{Flowchart of the Rosetta Stone end-to-end framework: Scheme of the comparison between observations and simulations by means of the production of realistic synthetic observations.
        Real data and inherent information are highlighted in green, synthetic data (RMHD simulations, radiative transfer, synthetic observations) in purple, and the steps of the analysis performed on both real and synthetic maps in orange.
        In the lighter-shaded boxes we specify how the building blocks of the Rosetta Stone framework have been adapted in this work for the SQUALO project (\citealt{2023MNRAS.520.2306T}) science case.}
        \label{fig:flow_chart}
    \end{figure*}
    While several studies in the literature employ post-processing routines that incorporate customized radiative transfer models and observational feature integration, they primarily address either the physics of isolated cores and the disk scales (e.g., \citealt{2007ApJ...665..478K}, \citealt{2012A&A...545A..98C}, 
    \citealt{2015MNRAS.453L..73D},
    \citealt{2019MNRAS.487.1248S}, \citealt{2022A&A...665A..25C},
    \citealt{2022FrASS...9.9223M},
    \citealt{2024A&A...685A..67R}, \citealt{2024A&A...684A..36T}) or the large-scale interstellar medium (e.g., \citealt{2014MNRAS.441.1628S}, \citealt{2015MNRAS.447.2144D}, \citealt{2018MNRAS.475.2383S}, \citealt{2019A&A...629A..63J}).
    A self-consistent and systematic comparison between actual observations and dust continuum emission synthetic observations, which could provide a quantitative characterization of how the initial conditions of clumps and environment affect the observed clump fragmentation properties, is currently missing in this panorama.
    To address this problem, we developed the Rosetta Stone project\footnote{\url{https://www.the-rosetta-stone-project.eu}}: an end-to-end (simulation $\Leftrightarrow$ observations) framework designed to systematically produce realistic synthetic observations of high-mass clump fragmentation and compare them with real interferometric data. 

    In the present work, we provide a detailed description of the Rosetta Stone framework. 
    In Section \ref{Sect:Method} we present our method, and we focus on the strategy designed for the production of synthetic observations.
    In Section \ref{Sect:Results} we report on the physical properties of fragments identified in the synthetic fields.
    Section \ref{Sect:Analysis} is dedicated to the analysis of the fragmentation properties (fragmentation level, mass accretion onto the fragments, relative distances between the fragments), and the impact of the initial conditions used in the models.
    In Section \ref{Sect:Discussion} we compare evidence from synthetic observations with both models and actual observations.
    Finally, in Section \ref{Sect:Conclusions} we draw the most relevant conclusions, and we provide insights into future research directions.

\section{The Rosetta Stone project}\label{Sect:Method} 
    Below, we introduce the Rosetta Stone post-processing pipeline aimed at producing realistic synthetic observations (see Section \ref{Sect:Synthetic observations}) starting from the output of radiative magnetohydrodynamical (RMHD) simulations (see Section \ref{Sect:Set of simulations}).
    The pipeline is designed to be flexible and can already be adapted to mimic any continuum survey performed with ALMA, the Very Large Array (VLA), and the Northern Extended Millimetre Array (NOEMA) interferometers.    
    In Fig. \ref{fig:flow_chart}, we provide the flowchart of the Rosetta Stone end-to-end framework and its main building blocks, which are, respectively:
    \begin{enumerate}[i.]
        \item The reference observations - Our first case study takes as reference ALMA 1.3 mm continuum dust emission observations from the Star formation in QUiescent And Luminous Objects (SQUALO) survey by \cite{2023MNRAS.520.2306T}. The SQUALO sample consists of $13$ high-mass star-forming regions of the Galaxy at different evolutionary stages, and all have evidence of infall motions at the clump scales.
        \item The suite of RMHD simulations - We rely on a new set of 24 radiative magnetohydrodynamical simulations of high-mass clumps fragmentation produced by means of the RAMSES code by Lebreuilly et al. (hereafter Paper I).
        \item The radiative transfer - Our simulations are post-processed at Spitzer 24 $\mu$m, {\textit{Herschel}} and ALMA Band 6 wavelengths by means of the RADMC-3D code by Tung et al. (hereafter Paper II).
        \item The synthetic observations - After the radiative transfer, our simulations are post-processed by means of the CASA software (\citealt{2007ASPC..376..127M}) to mimic the SQUALO survey observing strategy (ACA + \texttt{12\,m} array) and data reduction. In this paper we focus on the systematic production of synthetic observations of continuum dust emission, coherently with the reference observations.
        \item The source extraction and photometry - We identify compact sources and we compute the photometry using the \textit{Hyper} software (\citealt{2015A&A...574A.119T}).
        \item The common method of analysis and the comparison of the real and synthetic datasets - We compare the fragmentation properties recovered from the analysis of the synthetic sources with those obtained from the analysis of ALMA observations in \cite{2023MNRAS.520.2306T}.
        \item The refinement of models and follow-up observations - The comparison of the two datasets allows us to better tune the initial conditions of the models and optimize the observational setup and strategy for follow-up observations.
    \end{enumerate}   
    
\subsection{Setup: Simulations and radiative transfer}\label{Sect:Set of simulations}
    The RMHD simulations of high-mass clump fragmentation have been newly developed within the framework of the Rosetta Stone project and are described in detail in Paper I. 
    Specifically, we rely on a first suite of simulations labeled RS1.0, together with a supplementary collection of models used to explore the effect of different numerical prescriptions and to tune the fiducial setup.
    The RS1.0 suite is part of a planned series of simulations specifically tailored to investigate the formation of high-mass protoclusters at the sub-parsec scales.\\ 
    \indent Our simulations solve the three-dimensional equations for the conservation of gas mass, momentum, total energy, and radiation total energy, as well as the evolution of the magnetic field under the assumption of ideal MHD. 
    In these models, which are computed by means of the adaptive mesh refinement finite-volume code RAMSES (\citealt{2002A&A...385..337T}; \citealt{2006A&A...457..371F}), individual forming stars are represented by sink particles (\citealt{2014MNRAS.445.4015B}) that form self-consistently above a threshold of $n_{thre} = 10^9$ cm$^{-3}$.\\ 
    \indent Since our ultimate goal is to mimic the observations of real star-forming high-mass clumps in the early phases of their evolution, we simulate the collapse of uniform clouds initialized with values of mass, radius, and temperature chosen in agreement with observations.
    These values correspond to clump masses of 500 and 1000\,M$_\odot$, a radius of $\sim 0.4$\,pc (e.g., 
    \citealt{2014MNRAS.443.1555U}, \citealt{2017MNRAS.470.3882T},  \citealt{2021MNRAS.504.2742E}), and a temperature of 10\,K, as suggested by the peak temperature derived for prestellar clumps in the Galaxy (e.g., \citealt{2021MNRAS.504.2742E}). 
    To also reproduce a realistic clump environment, we initialize our models including self-gravity, stellar radiation due to mass accretion onto the sink particles, turbulence, and magnetic fields.
    In particular, we explore two different initial levels of turbulence quantified by the Mach number:
    \begin{equation}
        \mathcal{M}=\frac{v}{c_s}\,,
    \end{equation} 
    where $v$ is the gas speed, and $c_s$ is the speed of sound in the medium. 
    According to the results from \citet
    {2017MNRAS.470.3882T}, we set $\mathcal{M}$ to $7$ and 10, respectively.
    These values are also supported by the analysis of turbulence in clumps performed by e.g., \cite{2010A&A...515A..42R}, and \cite{2024A&A...686A.155C}.

    Given our limited understanding of the magnetic field impact on fragmentation (from observations alone), we explore three potential scenarios: a quasi-hydrodynamical one, a moderately magnetized one, and a more magnetized one.
    The clump magnetization of each scenario is quantified by the normalized parameter mass-over-flux to critical-mass-over-flux ratio (also known as normalized mass-to-flux ratio):
    \begin{equation}
        \mu = \biggl(\dfrac{M_0}{\phi}\biggr) \bigg{/}\biggl(\dfrac{M}{\phi}\biggr)_c \, ,
    \end{equation} 
    where $M_0$ corresponds to the mass of the clump, and $(M/\phi)_c=$\,$(0.53/\pi)\sqrt{5/G}$ corresponds to the critical mass-to-flux ratio at which the clump is magnetically stabilized against gravitational collapse (\citealt{1976ApJ...210..326M}).
    The stronger the magnetic field, the lower is $\mu$ for a clump of fixed mass.
    We set $\mu$ to $\sim3$, 10 and 100, respectively (Paper I). 
    As initial condition, the magnetic field is uniform and oriented along the $z$-axis.

    As anticipated above, this first suite of simulations has been developed including feedback in the form of radiation from luminosity accretion.
    This approach enables us to isolate the individual contributions and focus on the impact of turbulence and magnetic fields on clump fragmentation properties. 
    Moreover, the computational cost savings achieved through this choice of feedback mechanisms allow us to explore a significantly broader range of parameters (Paper I).
    Outflows and H{\sc ii} regions are planned to be incorporated in a future set of Rosetta Stone models.

\begin{figure*}
        \centering
        \includegraphics[width=\textwidth]{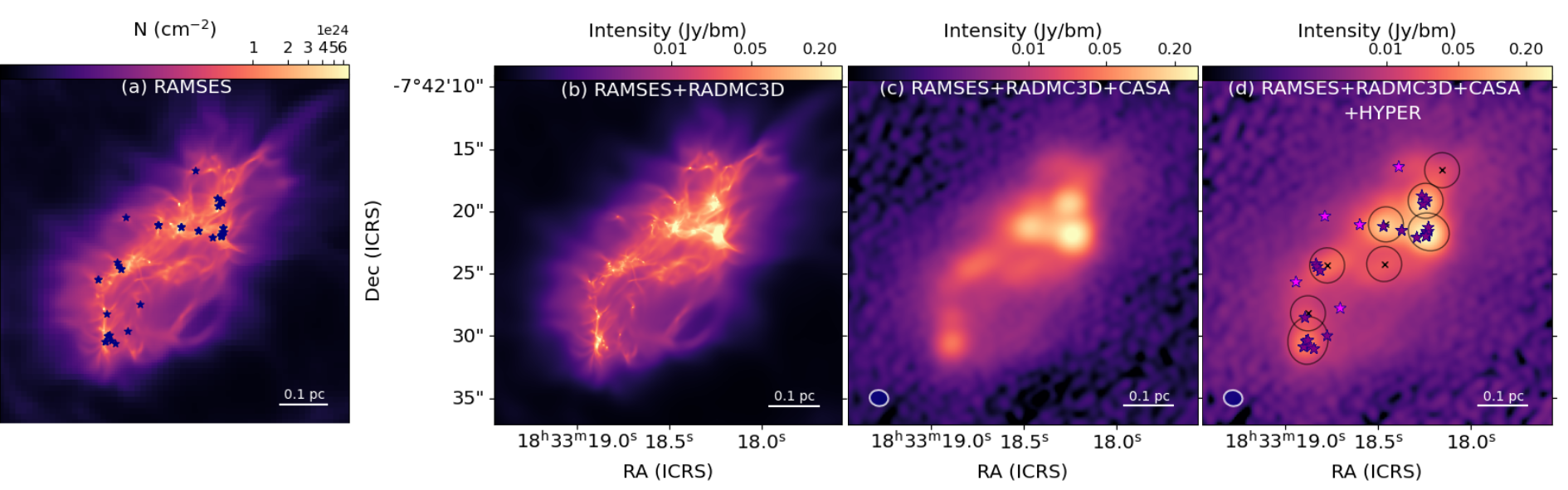}
        \caption{Example of the post-processing routine on a snapshot of clump collapse. This snapshot, belonging to the \textit{M1000\_$\mu$10\_$\mathcal{M}$7\_S2} realization,  corresponds to an intermediate level of SFE ($\sim15\%$). The displayed projection corresponds to the $x$-axis, i.e., the $y-z$ plane. Left to right:
        a)\,column density map obtained projecting the RAMSES volume density cube. The blue stars mark the projected positions of the sink particles; 
        b)\,intensity map after the RADMC-3D computation of the radiative transfer at 1.3\,mm; 
        c) synthetic intensity map after the post processing performed with the CASA software; 
        d) source extraction and photometry performed by means of the \textit{Hyper} software. The peaks identification is performed on the synthetic map in (c). Then, source photometry is computed on the PB corrected map, which takes into account the synthetic beam obtained by the combination of synthetic observations with both the \texttt{7\,m} and \texttt{12\,m} arrays. The black crosses and ellipses mark the centroids and the contours of the 8 identified fragments.
        Sink particles with and without a fragment counterpart are shown as purple and magenta stars, respectively.
        In each map we show the linear scale corresponding to $0.1$ pc, and in the steps involving CASA software we also show the 1.29$^{\prime\prime}\times$1.51$^{\prime\prime}$ synthetic beam footprint.}
        \label{fig:postprocessing_sequence}
    \end{figure*}

    With fixed physical initial conditions, we let each model evolve in time with two different turbulent seeds, $1$ and $2$ respectively, to account for  the stochastic fluctuations in the observed fragmentation process.
    We also test different time-step samplings across the various realizations, resulting in a varying number of saved time steps per realization. 
    A larger collection of time steps is available for the realizations initialized with Seed $2$.
    Each computational time step is associated with a specific time in years and a sink formation efficiency (SFE) computed as:
    \begin{equation}\label{eq:SFE}
        \text{SFE} = \frac{M_{sinks}}{(M_{gas} + M_{sinks})}\,,
    \end{equation}
    where $M_{sinks}=\sum_{\mathrm{sinks}} M_{\mathrm{sink}}$ and $(M_{gas} + M_{sinks})$ is the initial mass in the simulated box.
    As showed in Paper II, the SFE is the most reliable parameter for defining the evolutionary stage of the clumps, providing a direct comparison with observational parameters such as $L/M$. 
    To sample an evolution range of $\sim10^5\,$years from the formation of the first sink particle, we stop the calculations at a SFE of $\sim15\%$ and $\sim30\%$ for the \textit{M500}\,\footnote{From now on, we refer to groups of realizations that share a common initial parameter. 
    For example, we use \textit{M500} for the realizations starting with a 500 M$_\odot$ clump. 
    The same criterion is used to indicate a specific model, as in \textit{M1000\_$\mu$10\_$\mathcal{M}$7\_S2}, which is the representative model in Fig. \ref{fig:postprocessing_sequence}.} and \textit{M1000} realizations, respectively (Paper I).
    
    For each time step, three orthogonal projections are analyzed. 
    These are fixed for all realizations, and correspond to actual projections in the sky along the $x$, $y$, and $z$ axes.  
    Since our observational vantage point is fixed in real observations, such a study is of crucial relevance to quantify the impact of the line-of-sight projection effects on the analysis of the clump fragmentation properties. 
    In total, we rely on a set of 24 available combinations of initial conditions (see Table \ref{tab:initial_conditions}), corresponding to a statistically significant sample of 244 data cubes and 732 2D fields.
    For each data cube, we have access to the density, velocity, and B 
    fields, and to the gas temperature.
    For this first work, we focus on the 2D information relative to the projected density fields and projected temperatures.
    The column density maps also contain information about the projected positions of the sink particles as shown in Fig. \ref{fig:postprocessing_sequence} (a).    

    \begin{table}
        \centering
        \caption{Initial conditions for the RS1.0 simulations.}
        \begin{tabular}{cc}
        \hline
        \addlinespace[2.5pt]
        Parameter & Available initial conditions \\
        \addlinespace[2.5pt]
        \hline
        \addlinespace[2.5pt]
            Seed & [1, 2]\\
            \addlinespace[2.5pt]
            $M$ (M$_\odot$) & [500, 1000]\\
            \addlinespace[2.5pt]
            $R$ (pc) & [0.383]\\
            \addlinespace[2.5pt]
            $\mu$ & [3, 10, 100]\\
            \addlinespace[2.5pt]
            $\mathcal{M}$ & [7, 10]  \\
            \addlinespace[2.5pt]
            \hline
        \end{tabular}
        \tablefoot{Col. 1: Parameters to be set (top to bottom: turbulent seed, mass, radius, mass-to-flux ratio, and Mach number). Col. 2: Available initial conditions for each parameter.}
        \label{tab:initial_conditions}
    \end{table}
    
    Following the philosophy of the Rosetta Stone project, performing a comparison between this set of simulations and our reference observations requires two additional steps from our end-to-end pipeline.
    These consist in post-processing physical quantities from the simulations into observable radiation and 
    incorporating observational features into the simulated data to generate realistic synthetic observations, respectively.
    At this stage, we have to choose which set of observations we want to compare our models to.
    This is fundamental to replicate the emission recovered at a given wavelength, for a simulated field placed at a given heliocentric distance, as if it was observed with a specific facility.\\
    \indent
    In this paper, we employ as reference the ALMA 1.3\,mm continuum dust emission observations from the SQUALO project by \cite{2023MNRAS.520.2306T}.
    Clump and fragmentation properties from the SQUALO survey that are useful for the analysis carried out in this paper are reported in Appendix \ref{SQUALO}.
    In particular, we reproduce the observational strategy used to observe the target source HIGALBM24.0116+0.4897 (HG24), on which forthcoming studies from the SQUALO survey are focusing (\textcolor{cobalt}{Jones et al. in prep.}).
    The source is located at a heliocentric distance of 5.2 kpc (\citealt{2021A&A...646A..74M}). At the SQUALO angular resolution of $\sim1.4^{\prime\prime}$, we are therefore able to probe the fragmentation process at 
    $\sim7000$ AU scales.
    We emphasize that our post-processing pipeline is highly flexible and can be adapted to different surveys. 
    In a forthcoming paper we will adapt our pipeline to reproduce the observing strategy of the ALMAGAL survey (\citealt{2025A&A...696A.149M}) and discuss clump fragmentation at $\sim2000$ AU scales.\\
    \indent As a first step in the post-processing, we run the previously saved snapshots of our simulations through the radiative transfer code RADMC-3D (\citealt{2012ascl.soft02015D}).
    This includes the physics of photon emission and absorption, coupled with a ray-tracing approach:
    as UV photon packages move through the grid they get absorbed by dust and re-emitted in a new direction, with a new wavelength, both derived according to the prescription by \cite{2001ApJ...554..615B}.
    To initialize the radiative transfer, we make use of the distribution of sink particles in the 3D-space (volume density cube) and the gas temperature cube ought to derive the dust temperature using a thermal Monte Carlo (MC) computation (Paper II).
    Once the sources of radiation are identified, the total luminosity is divided into photon packages. 
    We refer to Paper II for a detailed description of the radiative transfer method.\\
    \indent We focus on the radiative post-processed dust continuum emission maps at 1.3\,mm as the one displayed in Fig. \ref{fig:postprocessing_sequence} (b).
    The flux is here computed with the same opacity assumption in use for the reference observation, i.e. $k_{1.3}=0.005\,\mathrm{g}\,\mathrm{cm}^{-2}$ (\citealt{1993A&A...279..577P}), that already includes a gas-to-dust mass ratio of 100.
    In Paper II, we also post-process the available simulations at Spitzer and Herschel wavelengths (24 $\mu$m and from 70 to 500 $\mu$m, respectively) to derive an estimate of the clump-averaged bolometric luminosity over mass ratio $L/M$, that \cite{2016ApJ...826L...8M} demonstrated to be a reliable evolutionary parameter.
    Specifically, Paper II establishes the quantitative relationship between the timing in SFE from the simulations in Paper I and the observational $L/M$ parameter:
    \begin{equation}
        \log(L/M) = 1.20_{-0.02}^{+0.02} \log(\text{SFE}) +3.28_{-0.03}^{+0.03}\,.
    \end{equation}

    We use this result in Section \ref{Sect:Analysis} and in Section \ref{Sect:Discussion} to investigate if and how the global evolution of our clumps - as seen through the eyes of an observer - correlates with fragmentation properties.
    The conversion from a numerical to an observationally-defined metric of clump evolution allows us to compare the fragmentation properties recovered in \cite{2023MNRAS.520.2306T} with the ones from synthetic clumps.  

\subsection{Synthetic observations}\label{Sect:Synthetic observations}  
    
    Actual observations result from the combination of real sky emission with the response of the imaging system, a process that cannot be replicated by the raw simulations.
    Effects as the smoothing caused by a finite beam size and the filtering of small or large-scale structures caused by interferometric observations, are examples of possible biases that the radiative transfer computation alone does not address.
    To mimic the features of the SQUALO project ALMA observations, we post-process our data using the CASA software (\citealt{2007ASPC..376..127M}), version 6.5.4. 

\subsubsection{Creating synthetic observations with the CASA software: Main challenges}
    The CASA software allows the creation of ALMA-like datasets starting from a sky model and incorporate both atmospheric and instrumental effects.
    For a given simulated source,  visibilities\footnote{Interferometers measure visibilities, a quantity obtained from the interference pattern formed by combining signals from different antennas.} can be computed using the  \texttt{simalma}\footnote{\url{https://casadocs.readthedocs.io/en/v6.5.4/api/tt/casatasks.simulation.simalma.html?highlight=simalma}} task for ideal ALMA interferometer arrays setups.
    However, during actual observing sessions the configuration of antennas may not match the ideal setup: some antennas might be out of use or poorly calibrated.
    Together with the elevation of the source, which often does neither reach the zenith, nor its culmination point, this causes the telescope beam to be more elongated than in the ideal case.
    Additionally, weather conditions might affect the data quality within the required integration time.  
    These factors introduce layers of complexity that must be considered to produce accurate synthetic observations, and can be included by tuning the CASA task \texttt{simobserve}\footnote{\url{https://casadocs.readthedocs.io/en/v6.5.4/api/tt/casatasks.simulation.simobserve.html?highlight=simobserve}}.
    
\subsubsection{Rosetta Stone (SQUALO) pipeline} 
    
    To address the aforementioned aspects and reproduce the observing conditions in the most accurate way possible, we make use of actual measurement sets and observations reports available for the SQUALO observations.
    Specifically, the archival data for the 2018.1.00443.S project (ALMA Cycle 6) include a single observing block on the \texttt{12\,m} array (C43-1 configuration), and five observing blocks on the \texttt{7\,m} array for the HG24 reference source.
    We used the observations reports to recover the following:
    \begin{itemize}
        \item The frequency of the central channel: 231.1 GHz, that is the 1.3 mm wavelength used for the radiative transfer;
        \item The fiducial antenna configurations, made up of $10$ to $12$ antennas for the execution blocks on the \texttt{7\,m} array, and $47$ antennas for the \texttt{12\,m} array; 
        \item The time on source, with $\sim470$ s of integration for each block on the \texttt{7\,m} array, and $\sim360$ s for the \texttt{12\,m} array;
        \item The time off source due to calibration necessity, which further splits each observing block into four intervals;
        \item The observing conditions, with a mean precipitable water vapor (PWV) of $\sim1.8$ and $\sim2.9$ mm computed over the observing blocks in the two different configurations;
        \item The (average) source elevation, covering a range from $\sim45^\circ$ to $\sim72^\circ$ in the different observing blocks.
    \end{itemize}

    Taking into account the details of each execution block, we used the CASA task \texttt{simobserve} to create synthetic measurement sets of the 1.3 mm sky models produced in Paper II, including the corresponding $uv$ coverage and visibilities.
    Then, the joint deconvolution with cleaning and masking was performed interactively with the task \texttt{tclean}\footnote{\url{https://casadocs.readthedocs.io/en/v6.5.4/api/tt/casatasks.imaging.tclean.html?highlight=tclean}} with the same pipeline settings that were applied in the imaging of the reference SQUALO source, namely:
    \begin{itemize}
        \item Spectral definition mode: Mfs, to get continuum imaging (Stokes I) with only one output image channel.
        \item Gridding options: mosaic, to resample visibilities onto a regular $uv$-grid with azimuthally symmetric beams without side lobes.
        \item Minor cycle algorithm (deconvolver): Multiscale with scales\,=\,[0,\,7,\,21], to look for point-like sources and multi-scale (multiples of the beam in pixels, corresponding to the beam and three times the beam) components left in the residual image to add them to the model image.
        \item Weighting scheme: Briggs with robustness parameter of 0.5. 
        We use a flexible weighting scheme based on the signal-to-noise ratio of the MeasurementSets and on the noise threshold. In \cite{2023MNRAS.520.2306T} it was possible to independently tune the cleaning noise threshold for the $13$ sources, here the threshold is automatically set according to the source evolutionary stage.
        \item Type of mask for deconvolution: Auto-multithresh, to automatically use multiple thresholds for deconvolution.
        \item Cell size: 0.2$^{\prime\prime}$, to sample our 1.29$^{\prime\prime}\times$1.51$^{\prime\prime}$ synthetic beam with $\sim 7$ pixels. With this setup, we reach a linear resolution of $\sim 7000$ AU (or $\sim0.036$\,pc) at the reference heliocentric distance of 5.2 kpc.
    \end{itemize}
    
    Cleaned images with and without primary beam (PB) correction, as the ones displayed in Fig. \ref{fig:postprocessing_sequence} (c) and (d) respectively, are produced along with the residual maps and the sources masks. 
    In the PB corrected maps, we recover from 80\% to 95\% of the flux present in the RADMC-3D projected maps as a function of clump evolution.
    Specifically, the interferometric filtering of the extended emission has a larger impact toward earlier stages of clump collapse.
    On the other hand, the flux retrieval is more effective toward advanced stages of clump evolution where most of the emission comes from the compact structures. 
    In Appendix\,\ref{Sect:consistency_checks}, we report on further consistency checks performed on the recovered r.m.s..
    In Appendix\,\ref{Sec:Pipelines comparison}, we address the differences between the results obtained with the Rosetta Stone pipeline and those obtained with \texttt{simalma}.
 
\subsection{Source extraction and photometry}\label{Sect:Source extraction and photometry}
    As for the previous post-processing steps, we want the source extraction and photometry to match that performed on the reference observations as much as possible.
    This approach further reduces the chance of introducing additional biases when comparing observations to simulations.
    Therefore, the source extraction is performed by using the extraction code \textit{Hyper}\footnote{The latest Python-based version of the \textit{Hyper} code is available at: \url{https://github.com/Alessio-Traficante/hyper-py}.} (\citealt{2015A&A...574A.119T}), which went through a fine-tuning procedure by \cite{2023MNRAS.520.2306T}.
    \textit{Hyper} has been originally designed to account for a multi-wavelength source identification and photometry of point-like and compact (elliptical) sources in regions (including crowded regions) where there is significant and variable background emission, and it is thus an optimal tool to use on interferometric observations of clusters formation.\\
    \indent Even though our strategy involves performing the source extraction as on the reference observations, we have also conducted an independent tuning of the software parameters.
    We have accounted for different detection thresholds, polynomials for background modeling, source sizes and shapes (defined by centroid position, position angle, pixel weighting).
    In agreement with the settings used in the SQUALO project, we considered as compact sources all identified objects with semi-axes\footnote{Semi-minor/major axes correspond to the \textit{Hyper} parameters full width at half maximum: FWHM\_MIN and FWHM\_MAX, respectively.} between one and two times the telescope beam size, having a maximum aspect ratio of 1.5.
    We modeled the background with polynomials up to the second order, and we performed the peak identification on the PB non-corrected images at three times above the locally estimated r.m.s., to then compute the photometry on the relative PB corrected images, as shown in Fig. \ref{fig:postprocessing_sequence} (d).\\
    \indent For each map in our sample of 732, we used \textit{Hyper} to estimate the fragment and field parameters.
    Specifically, we derived the number of identified fragments and the coordinates of their centroids. 
    We also determined the semi-minor and semi-major axes, along with the position angle, to characterize the region where the aperture photometry is performed.
    Within this same region, we measured the peak of emission, and the integrated flux.
    Additionally, we assessed the local background level for each fragment and the r.m.s. level characteristic of each field.\\
    \indent In Section \ref{Sect:Results}, Section \ref{Sect:Analysis}, and Section \ref{Sect:Comparison with numerical models} we report the results from the realizations with turbulent seed $2$, for which we sampled the clump evolution at a higher cadence obtaining a sample of 393 maps.
    Results from seed $1$ are reported in Appendix \ref{Sect: Seed1 fragmentation properties}.
    In Section \ref{Sect:Comparison with observations}, we use all the 732 maps to reconstruct the parameter space covered by the whole RS1.0 collection of synthetic observations.
        
\section{Results}\label{Sect:Results}
    Based on the criteria provided in Section \ref{Sect:Source extraction and photometry}, we are able to identify (and visually validate) 2749 fragments across the 393 maps under analysis.
    In the following, we report on the physical properties of the identified sample, with specific focus on the mass distribution and the associated uncertainties.

\subsection{Physical properties of the fragments}\label{Sect:Radius and mass of the fragments}
    
    The sample of validated fragments is characterized by radii in the range $0.036\leq R_f \leq 0.071$ pc, computed as the geometrical mean of the two FWHMs estimated by \textit{Hyper}.
    About $65\%$ of these fragments are symmetrical, of the remaining $\sim35\%$, $\sim20\%$ have an aspect ratio between $1$ and $1.25$, while $\sim15\%$ have an aspect ratio between $1.25$ and $1.5$ (the maximum aspect ratio allowed).
    
    The distribution of mass of the fragments was obtained by converting fluxes into masses as follows (\citealt{1983QJRAS..24..267H}):
    \begin{equation}\label{eq:mass_f}
        M_f=\dfrac{D^2 S_{1.3}}{B_{1.3}(T_f) \kappa_{1.3}}\,,
    \end{equation}
    where D is the distance of the reference source, $S_{1.3}$ the integrated flux at 1.3 mm estimated by \textit{Hyper} for each fragment, $B_{1.3}(T_f)$ is the Planck function computed at 1.3 mm at the chosen temperature $T_f$ for the dust envelope of each fragment, and $\kappa_{1.3}$ the dust absorption coefficient at 1.3 mm.
        
    As in \cite{2023MNRAS.520.2306T}, we adopt the single temperature model to derive the mass of the fragments associated with the same parent clump.
    We first assign a temperature to each clump based on the $L/M$ parameter, and then we compute the mass of the fragments therein.
    Thanks to the Rosetta Stone framework, we are now able to assign a value of $L/M$ to each individual time step on the basis of the associated SFE (see Section \ref{Sect:Method} and Paper II).
    The relative mass uncertainties due to the possible error on the temperature estimate is computed via 1000 MC simulations of fragments temperature, which let the temperature vary within fixed ranges.
    The $L/M$ ranges, the reference temperatures, and the relative temperature ranges are provided in Table \ref{tab:T_limits}.
    
    We do not expect the real dust temperature to significantly vary from the adopted one in the initial evolutionary stages.
    However, according to recent studies (e.g., \citealt{2025A&A...694A..24M}, \citealt{2025A&A...696A.151C}, \textcolor{cobalt}{Jones et al. in prep.}), a larger mismatch between the real temperatures and the ranges indicated in Table \ref{tab:T_limits} is expected in the more evolved stages.
    In this case, we face an additional limitation: the SQUALO clumps reached a maximum $L/M$ of 107 L$_\odot$/M$_\odot$, and therefore the $L/M$ range from 10 L$_\odot$/M$_\odot$ onward did not require further division into smaller sub-ranges. 
    As a consequence, the reference temperature of 40 K was not conceived to represent clumps with $L/M > 100$ L$_\odot$/M$_\odot$.
    
    Before assessing the clump fragmentation properties, we evaluate how effectively we can detect the faintest objects in each synthetic observation depending on the r.m.s. of the map. 
    As in \cite{2023MNRAS.520.2306T}, we consider the 1$\sigma$ value of the local r.m.s. as a lower limit to the peak emission that can be recovered in each map.
    Assuming that the 1$\sigma$ value corresponds to the peak flux emitted by a point-like source, 
    we compute the equivalent mass with Eq. \ref{eq:mass_f}. 
    From this procedure, we estimate a point-like source mass sensitivity in the range between $\sim0.02$ and $\sim5.4$\,M$_\odot$ across the whole sample.
    These two-order magnitude span depends on the temperature associated with each clump in a given evolutionary stage (see Table \ref{tab:T_limits}) and on the r.m.s. level characteristic of each map, which variations are characterized in Appendix \ref{Sect:consistency_checks}.
    \begin{figure}
        \centering
        \includegraphics[width=0.5\textwidth]{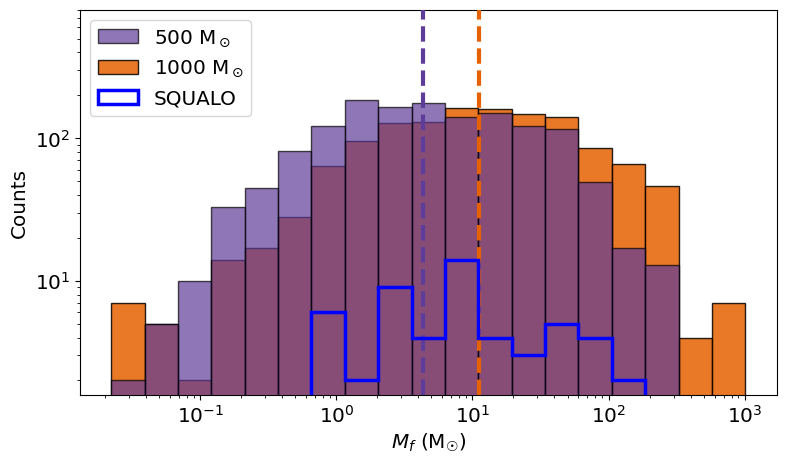}
        \caption{Mass distribution. The distributions relative to the \textit{M500} and \textit{M1000} samples are displayed in purple and orange, respectively. 
        The dashed vertical lines represent the median values of $4$ and $11$\,M$_\odot$.
        The mass distribution relative to the SQUALO sample is overlaid in blue.}
        \label{fig:mass_distr}
    \end{figure}

    \begin{table}
        \centering
        \caption{Reference temperatures based on clump $L/M$.}
        \begin{tabular}{ccc}
        \hline
        \addlinespace[2.5pt]
             $L/M$  & $T_f$ & $T_{lim}$\\
             (L$_\odot$/M$_\odot$) & (K)&  (K)\\
             \addlinespace[2.5pt]
        \hline\hline
        \addlinespace[2.5pt]
             $L$/$M < 1$  & 20 & 10 - 40 \\
             $1 < L$/$M < 10$ & 30 & 10 - 50 \\
             $L$/$M > 10$ &  40 & 20 - 60 \\
             \addlinespace[2.5pt]
             \hline
        \end{tabular}
        \tablefoot{Col. 1: $L/M$ ranges taken from \cite{2016ApJ...826L...8M}. Col.\,2:\,Reference temperature of the dust envelope used to estimate the fragments mass. Col. 3: Range of temperatures used in the MC simulations to estimate the mass uncertainties.}
        \label{tab:T_limits}
    \end{table}
    
    Using the single temperature assumption and Eq. \ref{eq:mass_f}, we obtain a mass estimate for each of the 2749 identified fragments.
    The overall sample spans a range between $\sim0.02$ and $\sim1010$\,M$_\odot$, as displayed in Fig. \ref{fig:mass_distr}.
    For the \textit{M500} realizations, 
    the range of mass covers structures from $\sim0.02$ to $\sim314$\,M$_\odot$. 
    For the \textit{M1000} realizations the range in mass is wider, from $\sim0.02$ to $\sim1010$\,M$_\odot$.
    We address this unrealistically high value in Section \ref{Sec:Temperature consistency} attributing it to the uncertainties on the clump temperatures and in Section \ref{Sect: mass recovery} where we also compare synthetically observed masses with the values obtained from the RAMSES column density maps.
    
    For each mass distribution we compute the median value, which corresponds to $\sim4$ and $\sim11$\,M$_\odot$, respectively.
    Taking into account the span in mass, and the median value of mass of each sample, we find that the initial reservoir has a role in the mass distribution of the object we recover: the most massive fragments are preferentially observed in the most massive clumps.
    The SQUALO distribution, ranging from 0.4 and 309\,M$_\odot$, falls exactly within the range spanned by both \textit{M500} and \textit{M1000}, as expected. 
    The minimum mass in the SQUALO distribution is one order of magnitude larger than the minimum of the synthetic distribution. 
    This can be a consequence of the range in mass sensitivity associated with the SQUALO sample, from 0.11 to 4.74\,M$_\odot$.
    While the synthetic clumps are all located at the same distance from the observer, SQUALO clumps are distributed across $13$ different distances, resulting in observations at varying resolutions. 
    We acknowledge that variations in clump distances are expected to introduce a slight bias on mass recovery.\\
    \indent We test the robustness of the results on the mass distribution using 1000 MC runs 
    in which a random temperature in the ranges presented in Table \ref{tab:T_limits} is assigned to each fragment.
    Following this strategy, the ranges of mass are updated: between $\sim0.01$ and $\sim724$\,M$_\odot$, and 
    between $\sim0.01$ and $\sim2338$\,M$_\odot$, for the \textit{M500} and \textit{M1000} realizations, respectively.
    Despite fluctuations in the mass range, the median values of the distributions remain stable.
    The current analysis demonstrates the importance of providing an accurate temperature estimate for our fragments in order to derive reliable mass measurements.
    For a more detailed discussion on the impact of temperature uncertainties on the mass estimation of our fragments, we refer to the following section.

\subsection{Temperature uncertainties}\label{Sec:Temperature uncertainties}

    Uncertainties on the temperatures of the dust envelope of the fragments - necessary to derive masses from both reference and synthetic observations - contribute significantly to the uncertainties associated with fragment mass estimates.
    The MC runs allow us to test the assumption of multiple temperatures for multiple fragments within the same parent clump.
    However, MC simulations are conducted on individual fields, without considering that we are observing subsequent time steps in the evolution of the same clumps. 
    This approach hampers our ability to track potential temperature trends throughout the clump evolution.
    In the case of synthetic observations, we can address these limitations using the projected temperature maps from the simulations.

    \subsubsection{Temperature consistency evaluation}\label{Sec:Temperature consistency}
    \begin{figure}
        \centering
        \includegraphics[width=0.5\textwidth]{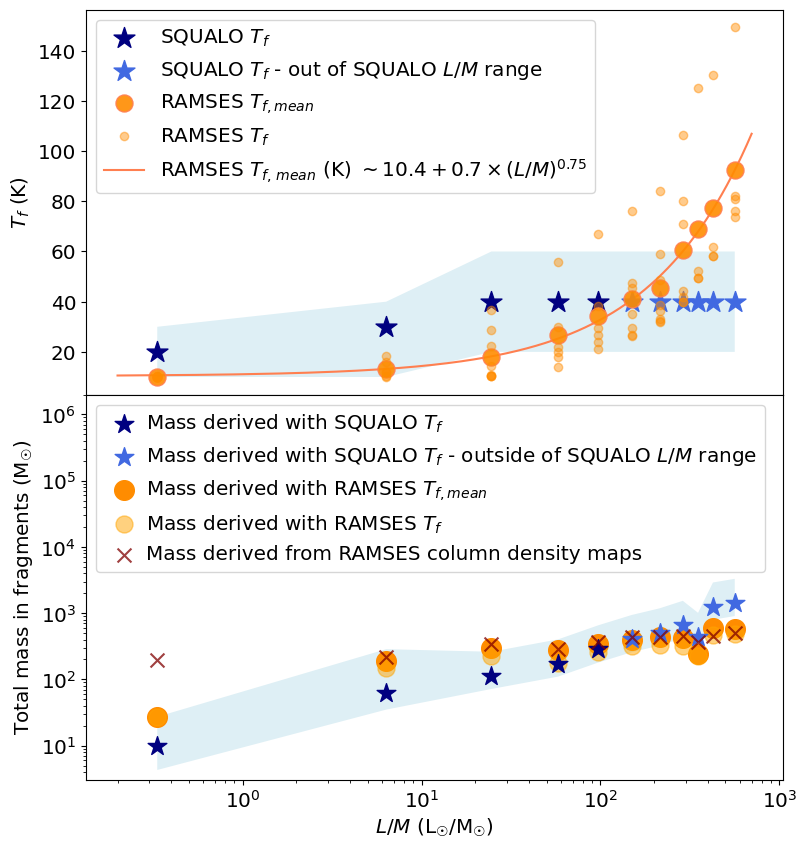}
        \caption{Evolution of fragments temperature (top) and mass (bottom) as a function of the clump $L/M$ for the \textit{M1000\_$\mu$10\_$\mathcal{M}$7\_S2} realization. The projection corresponds to the $x$-axis, i.e. the $y-z$ plane.
        In the top panel, dark blue stars mark SQUALO reference temperatures according to the evolutionary stage of the clump (see Table \ref{tab:T_limits}).
        Light blue stars mark the reference temperature in the range of $L/M$ not covered by the SQUALO sample.
        The shaded area corresponds to the temperature ranges used in the 1000 MC simulations of fragments temperature.
        Small and big orange dots mark, at each given time step, the RAMSES individual temperature of the identified fragments and the RAMSES mean temperature computed across all fragments in the same synthetic field, respectively.
        The fit of the RAMSES mean temperature as a function of clump $L/M$ is provided as a reference and corresponds to the orange solid line.
        In the bottom panel, stars and dots indicate mass estimates obtained using the  temperatures in the top panel (same color code).
        Red crosses indicate the mass estimates obtained by integrating the column density maps from the RAMSES simulations within the areas corresponding to the footprints of the identified fragments.}
        \label{fig:temperature_ev_ramses_vs_SQUALO}
    \end{figure}
    Using the known position of the peaks and the area covered by each fragment, we obtain the footprints of the fragments on the RAMSES projected temperature maps. 
    A mean fragment temperature can be then evaluated within these footprints.
    This kind of analysis, which cannot be reproduced on real observations, is useful to quantify the uncertainties on the assumed temperature values and ranges, along with its impact on mass estimates.\\
    \indent In Fig. \ref{fig:temperature_ev_ramses_vs_SQUALO} we compare the fragment temperatures and ranges adopted in \cite{2023MNRAS.520.2306T} with the fragment temperatures derived for the \textit{M1000\_$\mu$10\_$\mathcal{M}$7\_S2} realization (the projection in use corresponds to the $x$-axis, i.e. the $y-z$ plane).
    The simulations-based recovered temperatures gradually increase with the evolution of the clump, as predicted for the observations-based $T_f$.
    However, in absolute values, the latter does not always prove to have a good match with the temperatures computed from the simulations.
    By fitting the RAMSES mean temperature, computed across $\sim7000$\,AU fragments in the same synthetic field, we derive a power-law of:
    \begin{equation*}
        T_{f,\,mean}\,\textrm{(K)}\, \sim 10.4+0.7\times (L/M)^{0.75}\, \,.
    \end{equation*}
    This trend suggests that temperature at the fragment scales increases as a function of clump evolution faster than what observed at the clump scales, e.g., $T_{clump}$\,(K)\,$ \propto$\,$(L/M)^{0.22}$ for $L/M\geq$10\,L$_\odot$/M$_\odot$ (\citealt{2016ApJ...826L...8M}).\\
    \indent As observed in Fig. \ref{fig:temperature_ev_ramses_vs_SQUALO}, toward the evolutionary stages below $\sim30$ L$_\odot$/M$_\odot$, the fragment temperatures populate only the lower end of the SQUALO temperature ranges.
    This could be due to the absence of feedback mechanisms, as well as a general overestimation of the fragment temperatures in the SQUALO sample. 
    Clump temperatures in the early stages of collapse may also be strongly influenced by the external radiation field.
    Between $L/M$ $\sim30\,$L$_\odot$/M$_\odot$ and $\sim300\,$L$_\odot$/M$_\odot$, the prescription well represents both the mean temperature and the temperature range associated with the fragments in the field.
    Above $\sim300$ L$_\odot$/M$_\odot$, few fragments populate the upper end of the SQUALO temperature ranges.
    The others exceed the predicted maximum temperature (60 K), also reaching temperatures above 100 K.
    As aforementioned, the reference temperature of 40 K, as well as the $20 - 60$ K span, have not been conceived to represent clumps with $L/M > 100\,$L$_\odot$/M$_\odot$, which are outside the SQUALO $L/M$ range.
    The mismatch between the SQUALO assumption and the temperature estimates derived from the numerical simulation causes a systematical overestimation of the mass of the observed fragments in the most evolved clumps.\\
    \indent The results obtained from the Rosetta Stone pipeline are already noteworthy, particularly in helping observers define an appropriate range of temperatures for their $\sim7000$ AU resolution surveys. 
    A more detailed investigation of temperature and mass uncertainties arising from observational assumptions will be presented in future works, through the analysis of a new set of simulations that include jets and H{\sc ii} regions. 
    By incorporating the effects of these feedback mechanisms on local temperatures and, consequently, on the multiplicity and masses of the recovered fragments, we will be able to more realistically model the advanced stages of cluster formation.
    
\subsubsection{Mass consistency evaluation}\label{Sect: mass recovery}

    In this section we compare the variations in mass estimates caused by the different temperature assumptions with the masses obtained by integrating the column density maps from simulations.
    This analysis allows us to quantify to what extent the synthetically observed fluxes yield to outcomes that are consistent with the density distribution observed in the numerical models (Paper I).
    In particular, in Fig. \ref{fig:temperature_ev_ramses_vs_SQUALO} (bottom panel) we compare the total mass in fragments obtained in four different cases: 
    i) using the SQUALO reference temperatures and temperature ranges; ii) using the mean temperature across fragments derived from the RAMSES temperature maps; iii) using the mean temperature of individual fragments derived from the RAMSES temperature maps; iv) integrating the column densities in the simulations (within the footprints corresponding to the identified fragments).\\
    \indent 
    Where the SQUALO temperature assumptions are comparable to the average temperatures in the 2D temperature maps from simulations (see Fig. \ref{fig:temperature_ev_ramses_vs_SQUALO}, top panel), we obtain consistent mass estimates across the four methods.
    However, this is not always the case, as observed in e.g., the first evolutionary step.
    In this case, the SQUALO temperature assumption, as the averaged RAMSES temperatures, do not reflect the actual RAMSES temperature along the line of sight causing the underestimation of the total mass in fragments.
    We observe a factor $\sim10$ of difference with respect to the mass derived from the column density.
    Analogous results are discussed in Paper II.\\
    \indent 
    Additionally, the SQUALO-like mass estimates in the latest evolutionary stages exhibit a factor $\sim2$ of difference with respect to the results obtained using RAMSES temperatures or column density maps.
    This mismatch occurs because the SQUALO assumptions were designed to cover a narrower range of $L/M$, ending at $\sim100$\,L$_\odot$/M$_\odot$.
    In the most extreme cases, because of the lower end of the assumed temperature range (see Fig. \ref{fig:temperature_ev_ramses_vs_SQUALO} and Table \ref{tab:T_limits}), the most evolved clumps are observed to have a total mass in fragments that exceeds the mass assigned as initial condition in the simulations (see Fig. \ref{fig:tot_mass_MC_vs_LoverM}).
    These results indicate that a finer prescription on the reference temperature (and associated temperature range) should be made to correctly characterize the observed evolved clumps, especially in the case of $L/M>300$\,L$_\odot$/M$_\odot$.
    
\section{Analysis}\label{Sect:Analysis}
    In this section, we report on the analysis of the statistical fragmentation properties derived from the synthetic observations as a function of the $L/M$ clump evolutionary parameter estimated in Paper II.
    Specifically, we investigate the level of fragmentation 
    in each synthetic map, the accretion of mass onto the fragments (total mass recovered in fragments with respect to the clump mass and evolutionary stage), and the relative (minimum and maximum) distances between the fragments.
    Specific focus is on the relative contributions of turbulence and magnetic fields, which cannot be directly studied using the 1.3 mm continuum data from \cite{2023MNRAS.520.2306T}. 
    We use the variations observed across the three available projections to provide an estimate of the uncertainties associated with the derived mean fragment properties at $\sim7000$ AU.

    Determining the most accurate estimate of fragmentation properties from synthetic maps is a challenging task, as demonstrated by the initial assessment of how well the masses can be recovered using observationally driven assumptions (see Section\,\ref{Sect: mass recovery}).
    A more robust characterization would require identifying three-dimensional, fragment-like structures directly within the simulations smoothed at $\sim7000$\,AU, which is currently beyond the scope of this work.
    Rather than ensuring $100\%$ consistency between simulated and synthetic quantities recovered at 1.3 mm, our primary goal is to compare the fragmentation properties from the post-processed maps with the SQUALO results within a self-consistent framework (see Section \ref{Sect:Comparison with observations}).
    To this end, we rely on the observationally driven assumptions presented in \cite{2023MNRAS.520.2306T}, even if this approach introduces some known biases.

\subsection{Fragmentation level}\label{Sect:Fragmentation level}
    
    \begin{figure*}
        \centering
        \includegraphics[width=\textwidth]{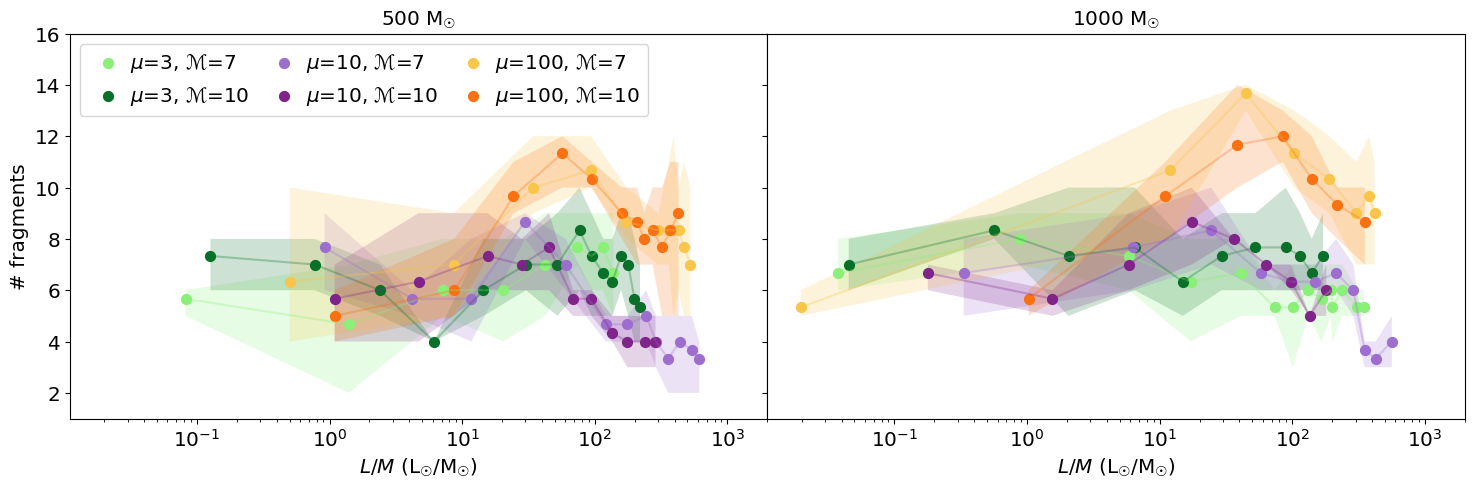}    
        \caption{Number of fragments identified in each clump as a function of the clump $L/M$. Left: Realizations with initial clump mass of 500 M$_\odot$. Right: Realizations with initial clump mass of 1000 M$_\odot$. Shaded areas illustrate the range in number of fragments recovered across the three projections, indicating the uncertainty in the distribution. The color code accounts for the different initial conditions of the realizations, with particular interest on the values assumed by the $\mu$ parameter: $\mu=3$ for the green dots, $\mu=10$ for the purple dots, and $\mu=100$ for the orange dots, respectively. }
        \label{fig:n_frag_vs_LoverM}
    \end{figure*}
    \begin{figure*}
        \centering        
        \includegraphics[width=\textwidth]{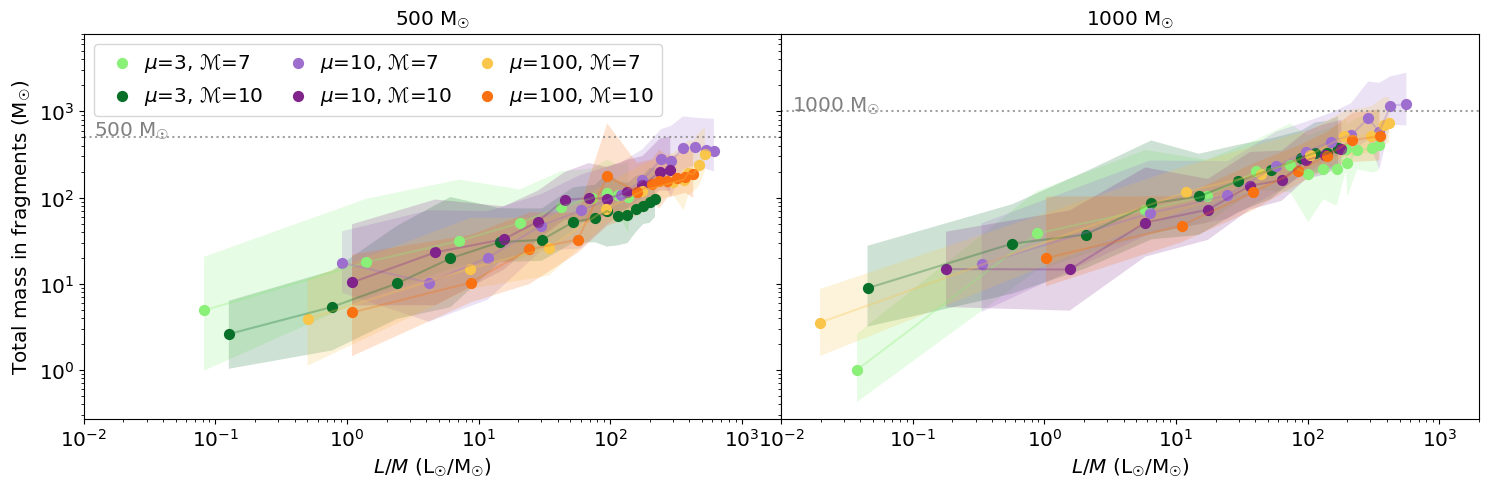}
        \caption{Total mass accreted onto the fragments as a function of the clump $L/M$.
        Left: Realizations with initial clump mass of 500 M$_\odot$. Right: Realizations with initial clump mass of 1000 M$_\odot$.
        The temperature of each fragment used to convert the fluxes into total mass (scatter plot) has been assigned using MC simulations, following the ranges in Table \ref{tab:T_limits}. The span in mass marked by the shaded areas accounts for stochastic variations due to 1000 different MC runs. The color code accounts for the different initial conditions of the realizations, as described in Fig. \ref{fig:n_frag_vs_LoverM}. The  horizontal dashed lines correspond to the initial condition on the clump mass: 500 and 1000 M$_\odot$, respectively.}        \label{fig:tot_mass_MC_vs_LoverM}
    \end{figure*}
    \begin{figure*}
        \centering
        \includegraphics[width=\textwidth]{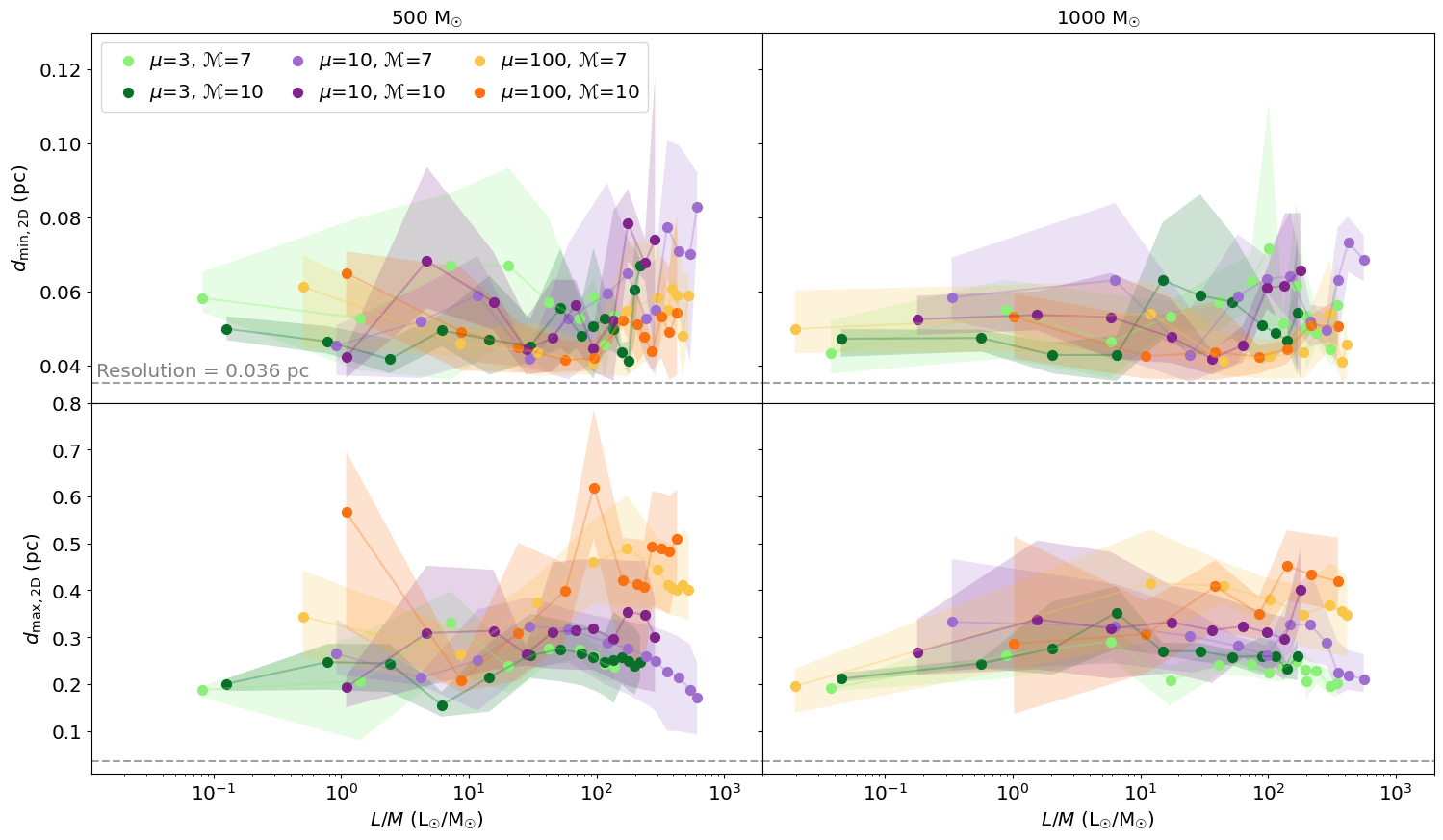}
        \caption{Minimum (top) and maximum (bottom) relative distance between the fragments as a function of the clump $L/M$. Left: Realizations with initial clump mass of 500 M$_\odot$. Right: Realizations with initial clump mass of 1000 M$_\odot$.
        Shaded areas illustrate the range in relative distance recovered across the three projections, indicating the uncertainty in the distribution.
        The color code accounts for the different initial conditions of the realizations, as described in Fig. \ref{fig:n_frag_vs_LoverM}. 
        The horizontal dashed line corresponds to the resolution limit of the synthetic observations, which is the minimum observable source separation.}
        \label{fig:mst_vs_LoverM}
    \end{figure*}  
    As the clumps collapse and sinks begin to form in the models, we can trace the emergence of fragments in the synthetic observations.
    In Fig. \ref{fig:n_frag_vs_LoverM}, we present the evolution of the number of identified fragments as a function of the $L/M$ evolutionary parameter. 
    This analysis traces the clump evolution for the available combinations of initial conditions, considering both the mean values and the spread observed across the three different projections. 
    For clarity, the plot is divided into two panels to separate the \textit{M500} and \textit{M1000} realizations.\\
    \indent The fragmentation levels we observe within individual fields range from $2$ to $14$ fragments, with an average of $\sim7$ fragments.
    Across the 393 maps, we do not find cases in which fragmentation is suppressed.
    On the contrary, our analysis indicates that fragmentation takes place within the clumps at every evolutionary stage, and that the number of fragments within each field is only weakly dependent on the initial mass of the clump.
    This suggests that physical factors beyond the initial mass are more relevant in determining the fragmentation patterns.\\
    \indent When examining the distribution of fragment multiplicities for individual realizations, we do not observe a clear, consistent trend with evolution.
    Instead, we note a general decrease in the number of fragments toward the more advanced phases of clump collapse. 
    This could be due to both pre-existing and newly formed fragments moving closer to the gravitational potential well(s) and merging as the clump evolves. 
    At $\sim 7000\,$AU resolution, what appears to be a merger of fragments, is not necessarily a physical merger.
    It is likely the result of multiple structures that may get closer than our resolution limit (see Section \ref{Sect:Minimum spanning tree and maximum distance}). 
    From this perspective, the global collapse of the clump appears to dominate the number of detectable fragments at $\sim 7000\,$AU resolution, especially in the latest evolutionary stages.\\
    \indent Comparing the level of fragmentation for the three available clump magnetization (i.e., $\mu=3$, 10, and 100)
    we observe that, for $L/M \gtrsim 20$ L$_\odot$/M$_\odot$, a larger number of fragments are preferentially formed when the magnetic parameter $\mu$ is set to the highest of the three explored values. 
    However, while magnetized and non-magnetized scenarios can be disentangled within the uncertainties, it is not possible to distinguish between the two magnetized realizations based solely on the number of identified fragments:
    multiplicity in the \textit{$\mu$3} realizations is higher than that in the \textit{$\mu$10} realizations during the later stages of the \textit{M500} clump collapse, while it is comparable in the \textit{M1000} case.\\
    \indent In addition to the influence of magnetic fields, we investigate the role of turbulence in driving the fragmentation properties of our synthetic clumps. 
    As shown in Fig. \ref{fig:n_frag_vs_LoverM}, the explored levels of turbulence do not appear to have a significant effect on the multiplicity of fragments.
    Realizations that differ only for the initial value of $\mathcal{M}$, are indeed comparable. 
    It is possible that the values chosen for the Mach number, $7$ and $10$, are too close to induce any significant variation in the properties of the structures identified at $\sim 7000\,$AU.\\
    \indent To further characterize potential relationships between the number of identified fragments and the initial conditions of clump collapse, we also use a suite of out-of-the-grid control models.
    Specifically, we investigate the impact of the initial volume density profile by setting a Bonnor-Ebert profile (realizations \textit{M1000\_$\mu3\_\mathcal{M}$7\_S2} and \textit{M1000\_$\mu10\_\mathcal{M}$7\_S2}, respectively), the impact of a lower level of turbulence by setting the initial Mach number equal to $3$ (realization \textit{M1000\_$\mu10\_\mathcal{M}3$\_S2}),  the impact of stochastic processes by including two additional seeds (realizations \textit{M1000\_$\mu$10\_$\mathcal{M}$7\_S3} and \textit{M1000\_$\mu$10\_$\mathcal{M}$7\_S4}, respectively). 
    Moreover, we evaluate the impact of the numerical prescription for the accretion luminosity coefficient $f_{\textrm{acc}}$, used to compute the efficiency of the accretion luminosity by testing different values of this parameter with respect to the reference value of 0.1 (for the \textit{M1000\_$\mu$10\_$\mathcal{M}$7\_S1} realization; Paper I). 
    Results from the analysis of these control models are reported in Appendix \ref{Appendix:ootg_models}.

\subsection{Accretion of mass onto the fragments}\label{Sect:Accretion of mass onto the fragments}

    After presenting the characteristic mass distribution of the fragments across all the available evolutionary stages of the parent clumps (see Fig. \ref{fig:mass_distr}), we now turn our attention to how the total mass recovered in fragments changes as a function of the $L/M$ evolutionary parameter. 
    This is done by calculating the mass of each fragment, assuming a single-temperature model (see Table \ref{tab:T_limits}), and then summing these values for each field. 
    In Fig. \ref{fig:tot_mass_MC_vs_LoverM}, we present the mean total mass accreted onto the fragments, which is calculated by considering the three different projections available at each time step.
    In this case, we use the 1000 MC simulations (see Section \ref{Sec:Temperature uncertainties}) to indicate the uncertainty on the distribution.\\
    \indent The results reveal that uncertainties in the temperature of the dust envelope surrounding the fragments contribute significantly to the uncertainties in the total mass estimate. 
    The variation in mass due to different temperature assumptions can lead to total mass estimates that are up to $\sim5$ times higher than the lowest value they can assume.
    As mentioned in Section \ref{Sec:Temperature consistency} (see also Section \ref{Sect: mass recovery}), the lower ends of the temperature ranges are responsible for overestimating the fragment masses, particularly in the more evolved stages of clump collapse.
    As a consequence, in most of the MC runs, the most evolved clumps happen to have a total mass in fragments that exceeds the mass assigned as initial conditions. \\
    \indent By examining the evolution of the total mass in fragments at $\sim 7000$ AU, we find that a continuous accretion process from the parent clumps to the fragments is taking place. 
    The seeds of star formation initially have low masses, often less than 1 M$_\odot$, and their final masses are progressively built up over time.
    Even relatively large fragments appear to be accreting mass from the parent clump.
    This provides clear evidence that the fragments do not evolve in isolation, but are instead fed by material from the parent clump which act as a mass reservoir.
    It is significant that this trend holds regardless of the different environmental initial conditions explored in the simulations. 

\subsection{Relative distance between the fragments}\label{Sect:Minimum spanning tree and maximum distance}

    Among the temperature-independent quantities of interest, we investigate the relative distances between the fragments as a function of the $L/M$ evolutionary parameter.
    Specifically, we focus on the minimum 2D distance ($d_{\text{min,2D}}$) between the fragments, already explored in \cite{2023MNRAS.520.2306T}, as well as on their maximum 2D distance ($d_{\text{max,2D}}$).
    Both quantities are derived from the distance matrix of the fragments, using the centroids of the fragments as edges to compute the relative distances.
    Together, they provide complementary information about the distribution of the fragments within the clumps at different evolutionary stages, potentially revealing important information about the mechanisms driving the approach or separation of the fragments during clump evolution.\\
    \indent In Fig. \ref{fig:mst_vs_LoverM}, we observe the oscillation in time of both $d_{\text{min,2D}}$ and $d_{\text{max,2D}}$. 
    The oscillation occurs regardless of the initial conditions of the simulations, suggesting that the observed behavior is either a general characteristic of the fragment dynamics at $\sim 7000$ AU scales or the result of a systematic observational bias, rather than being driven by specific starting configurations.
    The resolved fragments appear to be distributed rather evenly across the parent clumps at each evolutionary stage.
    However, this result might be strongly biased by the linear resolution of $\sim0.036$ pc, which is the minimum observable source separation determined by the combination of the synthetic beam and the reference distance of our sources. 
    We acknowledge that, in the case of rapid collapse, we would be unable to detect new structures forming at scales below the available resolution, or to track already identified structures that get closer than this minimum separation.\\
    \indent We find that slightly smaller mean values of $d_{\text{max,2D}}$ - reflecting a smaller separation between the most distant fragments - are preferentially observed toward evolved, magnetized clumps.   
    As in the case of the number of fragments (see Section \ref{Sect:Fragmentation level}), the evolutionary stage after which the realizations with different clump magnetization start to behave differently corresponds to $L/M\sim20$ L$_\odot$/M$_\odot$.
    Although we see a difference in the mean values of $d_{\text{max,2D}}$, this result is not always robust across the different projections.
    For the \textit{$\mu$3} realizations, the fluctuations associated with the three lines of sight are of similar magnitude to the variations introduced by different initial magnetic field conditions, causing its distribution to overlap with that of the non-magnetized scenario at the upper limit.\\
    \indent The different values observed for $d_{\text{min,2D}}$ and $d_{\text{max,2D}}$ may possibly arise from geometrical effects related to the combination between the size of the parent clumps and the number of fragments formed within them.
    In our case, it is likely that stronger magnetic fields are shaping the interstellar gas leading to more tightly bound configurations of the parent clumps and, consequently, to smaller $d_{\text{max,2D}}$. 
    However, the observed oscillations and the spread in the data caused by the line of sight effect limit our ability to definitively attribute the recovered fragment distribution to the clump magnetization alone. 

\section{Discussion}\label{Sect:Discussion}

    Building on the results and the analysis presented in Section \ref{Sect:Results} and Section \ref{Sect:Analysis}, we compare the synthetic observations produced at $\sim 7000$ AU resolution with the output of the simulations presented in Paper I.
    Furthermore, we compare results from our synthetic observations with actual observations, with particular focus on the fragmentation properties analyzed in the SQUALO project (\citealt{2023MNRAS.520.2306T}, see Table \ref{tab:SQUALO}).

\subsection{Comparison with numerical models}\label{Sect:Comparison with numerical models}

    The systematic production of synthetic observations of clumps undergoing the fragmentation process is a yet to be established approach (e.g., 
    \citealt{2018NewAR..82....1H},
    \citealt{2020SSRv..216...62R}), and to the best our knowledge, there are no existing examples of systematic comparisons of fragmentation properties between the synthetic observations and the numerical simulations from which they are derived.
    Such comparison serves a dual purpose: first, to assess whether the synthetic observations are reliable proxies for the physical outcomes of the underlying models, and second, to ensure that the numerical computation - native at a numerical resolution of $\sim40$ AU (Paper I) - accurately captures the key physical processes, even on larger scales.
    
\subsubsection{Fragments versus sinks multiplicity}\label{Sec:Comparison with number of sinks in the simulations}
    
    A substantial discrepancy exists between the multiplicity observed in the simulations and in the corresponding synthetic observations.
    The number of sink particles identified in the models can reach up to the order of 100 in the later stages of clump evolution, with this value being highly sensitive to the initial conditions of the clump collapse, particularly the clump magnetization (up to $\sim50$ sinks of difference across different clump magnetization; Paper I). 
    In contrast, the identification of fragments is strongly affected by the resolution limit covering a narrow range of observed fragment multiplicity ($2-14$), as discussed in Section \ref{Sect:Fragmentation level}, and supported by results from \cite{2018A&A...615A..94F} and \cite{2021A&A...653A.157L}.  
    
    To address the discrepancy in sinks/fragments multiplicity, we make use of the 3D positional information of the sink particles within the simulated volume to then examine the correspondence between the 2D projected positions of the sinks and the locations of the identified fragments in our synthetic observations. 
    This analysis allows us to evaluate the nature of the structures identified at $\sim7000$ AU and assess the extent to which the fragmentation properties at this scale reflect the underlying physical processes in the simulations.
    As illustrated in Fig. \ref{fig:postprocessing_sequence} (d), we distinguish different scenarios of fragment-sink(s) correspondence, with their respective statistics reported in Table \ref{tab:fragments_sinks_statistics}.
    \begin{table}[]
        \centering
        \caption{Number of fragments and sink particles.}
        
        \begin{tabular}{ccc}
        \hline
        \addlinespace[2.5pt]
             & \# & \% \\
        \addlinespace[2.5pt]
        \hline
        \addlinespace[2.5pt]
           Fragments  & 2749 & 100 \\
           Fragments w/ sinks & 2066 & $\sim75.2$ \\
           Fragments w/ one sink & 458 & $\sim16.7$\\
           Fragments w/ multiple sinks & 1608 & $\sim58.5$ \\
           Fragments w/out sinks & 683 & $\sim24.8$ \\
           \addlinespace[2.5pt]
           \hline
           \addlinespace[2.5pt]
           Sinks  & 16209 & 100\\
           Sinks w/ fragments & 12389 & $\sim76.4$\\
           Sinks w/out fragments & 3820 & $\sim23.6$\\
           \addlinespace[2.5pt]
        \hline
        \end{tabular}
        \tablefoot{Col. 1: Variables related to fragments and sink particles. Col.\,2:\,\,Absolute number. Col. 3: Percentage with respect to the total number of fragments and sink particles, respectively.}
        \label{tab:fragments_sinks_statistics}
    \end{table}
    Focusing on the fragments:

    \begin{itemize}
        \item 
        $\sim75$\% of the identified fragments have a correspondence with sink particles. 
        Only $\sim 17$\% of the total are associated with a single sink particle;
        \item $\sim25$\% of the identified fragments do not correspond to any sink particle.
    \end{itemize}

    The existence of multiple sinks within individual fragments supports a hierarchical and/or multi-layered fragmentation process, suggesting that fragmentation occurs across multiple scales simultaneously.
    Larger clumps break into smaller fragments, and within these fragments, further fragmentation can occur, leading to the formation of one or more sinks. 
    The differing sensitivity of fragment and sink multiplicity to the clump magnetization indicates that magnetic fields influence the fragmentation process at different scales, though with varying degrees of significance.\\
    \indent
    The case of identified fragments that do not correspond to any sink particle, and therefore are not actively forming stars at a given time, can be explained by the following scenarios: i)\,They correspond to overdensities that have not yet reached the volume density threshold required to form a sink particle but will reach it in the subsequent steps of clump evolution. ii) They correspond to overdensities that are transient and will not form a sink particle in the subsequent steps of clump evolution. Some of these will eventually be dispersed. iii) They appear as overdensities due to projection along the line of sight and/or are the artificial result of the convolution of diffuse structures with the beam of the telescope. \\
    \indent
    Focusing on the sink particles, we find that $\sim76$\% of them exhibit a corresponding fragment, while the remaining $\sim24$\% do not have a correspondence with the identified fragments.
    Our analysis reveals that this lack of correspondence never involves the most massive sink particle in each field. 
    It is also noteworthy that the absolute number of sink particles without a matching clump varies with the chosen value of $\mu$. 
    For the \textit{$\mu$100}, \textit{$\mu$10}, and \textit{$\mu$3} realizations, this goes up to $\sim 20 $, 13, and 8 sinks per field, respectively. 
    This result confirms our ability to best recover compact structures within magnetized clumps.
    Magnetic fields cause initially homogeneous clumps to collapse in dense and compact sheets and/or filaments (e.g., Paper I, \citealt{2019FrASS...6....5H}), reducing the presence of diffused matter more efficiently than in the quasi-hydrodynamical case.
    As a potential consequence, compact fragments that exceed the 3$\sigma$ threshold might be more effectively formed in the magnetized scenarios.\\
    \indent We will be able to address these specific aspects more effectively by closely tracking the evolution of individual fragments through the use of tracer particles, which we plan to include in a forthcoming suite of simulations.
    As of now, the number of fragments does not correspond to the number of sink particles due to limitations in both spatial resolution and sensitivity (some sinks may not be bright enough to be detected as fragments arising above the 3$\sigma$ level). 
    As a result, the increasing number of sink particles\footnote{In our prescription, the merging of sink particles is not allowed. Consequently, the number of sink particles can never decrease.} with the progressive evolution of individual clumps and the impact of clump magnetization - discussed in Paper I, and in line with the models by e.g., \cite{2011ApJ...729...72P}, \cite{2018A&A...615A..94F},  \cite{2022A&A...668A.147H} - are not equally well observed in the number of fragments at $\sim7000$ AU. 
    Beside the observational biases, this evidence might suggest that the physical mechanisms explored in the simulations could have different significance at the different physical scales.

\subsubsection{Comparison with sink formation efficiency from the simulations}

\begin{figure}
        \centering
        \includegraphics[width=0.5\textwidth]{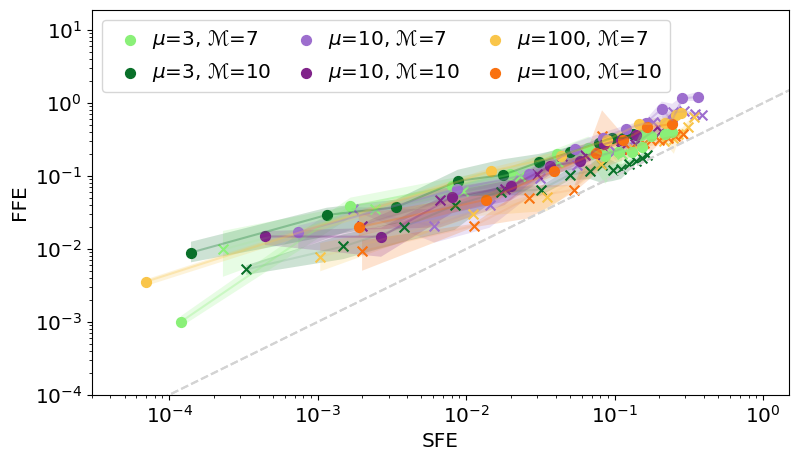}
        \caption{Fragment formation efficiency vs. sink formation efficiency. 
        The color code accounts for the different initial conditions of the realizations, as described in Fig. \ref{fig:n_frag_vs_LoverM}. 
        Specifically, the crosses and the points mark the \textit{M500} and \textit{M1000} realizations, respectively.
        Shaded areas illustrate the range in FFE recovered across the three projections, indicating the uncertainty in the distribution.
        The diagonal dashed line marks the equality between fragment formation efficiency and sink formation efficiency.
        Data points are expected to lay above the line. 
        Only if the sink particles were resolved, the fragment formation efficiency could align with the one measured for the sinks.
       }
        \label{fig:FFE_vs_SFE}
    \end{figure}

    As previously discussed, the SFE parameter, defined as the total mass recovered in sink particles with respect to the mass of the clump (see Eq. \ref{eq:SFE}), can be used in simulations as a proxy of the evolution. 
    In a similar manner, we can define a corresponding quantity at the spatial scales probed by the synthetic observations: the fragment formation efficiency (FFE). 
    The FFE is calculated by dividing the total mass recovered in fragments by the mass of their parent clump, offering a way to track the evolution of the clump at intermediate scales.\\
    \indent The comparison between the FFE and the SFE provides a valuable tool for interpreting the distribution of mass into compact structures at various physical scales. 
    By examining how much mass is condensed into $\sim 7000$ AU fragments and sink particles at different stages of clump evolution, we gain first insights into the hierarchical fragmentation process.
    As shown in Fig. \ref{fig:FFE_vs_SFE}, the FFE consistently exceeds the SFE at each evolutionary stage. 
    Fragments at $\sim 7000$ AU are more than the simple product of the convolution of one/more sink particles with the telescope beam.
    Rather, they are real, compact structures that are made up of gas which is more diffuse in comparison to the sinks.
    This evidence suggests that a significant fraction of the mass within the clump is condensed into fragments, and only a small fraction of it is ultimately converted into stars.\\
    \indent In particular, in the earliest stages of clump collapse, at SFE$\,\,\sim\,10^{-3}$, the majority of the gas locked into the fragments (FFE $\sim\,10^{-2}$) has not yet been converted into sinks.
    The difference in these initial efficiencies is further evidence in favor of a multi-layered fragmentation process, where small-scale sinks form within intermediate-scale fragments.
    With the clump evolution, the mass reservoir available to the fragments diminishes.
    On the contrary, the sinks accrete with high efficiency from the fragments, bringing the FFE vs. SFE relation closer to the one-to-one line toward the most advanced evolutionary stages.
    Thanks to the forthcoming RS2.0 suite of simulations, we will explore the same relation in the case of clumps that are formed within filaments.
    Such analysis will allow us to assess if a depletion in the clump reservoir is observed also when the parent clump can potentially accrete from a larger filament (in opposition to the isolated box).\\
    \indent In actual observations, the trend in accretion efficiency onto compact structures across scales shows a marked decrease, from values up to $\geq$40\% when estimated from clumps down to $\sim$5000\,AU structures, to $\leq$10\% when considering the efficiency from clumps down to the smallest fragments and cores\footnote{Here, we specifically define cores as units that will eventually form a single protostar or a small-multiplicity system.}, as outlined by \textcolor{cobalt}{Traficante et al., in prep.}.
    We will also test this result thanks to the production of high-resolution ($\sim 2000$ AU) synthetic observations.

\subsection{Comparison with observations}\label{Sect:Comparison with observations}
    
    Synthetic observations serve as a bridge between theory and actual observations, allowing for a self-consistent comparison in terms of common metrics and scales under analysis. 
    The following comparison with the reference observations further proves the predictive power of synthetic observations in the context of star formation.
    It also represents an important step in the scientific validation of both the RS1.0 simulations and their derived synthetic counterparts.\\
    \indent
    Previous studies have certainly benefited from direct comparisons between observations and numerical models (e.g., \citealt{2012ApJ...761..156F}, \citealt{2013A&A...555A.112P}, \citealt{2019ApJ...886..102S}, \citealt{2020A&A...644A..52A}, \citealt{2021A&A...645A.142A}, \citealt{2024ApJS..270....9X}), but additional layers of complexity must be taken into account when interpreting actual observations on the basis of numerical outputs.
    Specifically, observers must first understand what kinds of structures and fragmentation properties the models generate at the scales they are interested in, since these are likely to be different from the observed sinks properties (see Section \ref{Sect:Comparison with numerical models}).
    
\subsubsection{Fragmentation level}

    \begin{figure}
        \centering
        \includegraphics[width=0.5\textwidth]{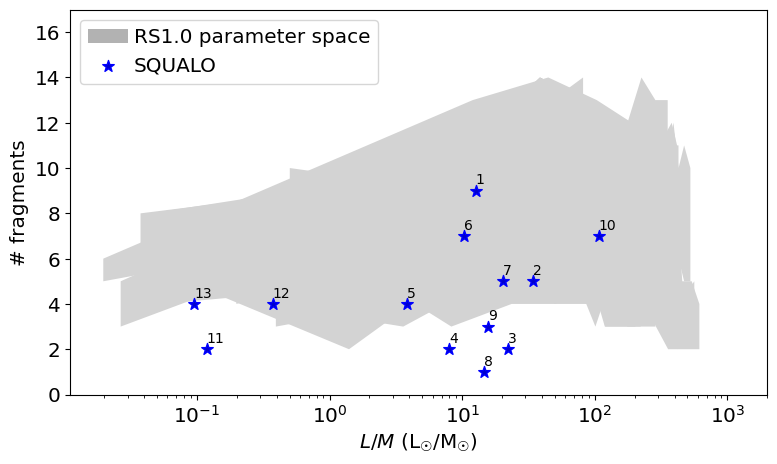}
        \caption{Number of fragments identified in each clump as a function of the clump $L/M$ and comparison with the SQUALO sample. 
        All the combinations of parameters presented in Table \ref{tab:initial_conditions} are included in the gray area. 
        The stars mark the fragmentation level of the $13$ clumps of the SQUALO sample. 
        }
        \label{fig:parameterspace_1}
    \end{figure}
    
    Based on the results from synthetic observations alone, the level of fragmentation observed in clumps at a spatial resolution of $\sim 7000$ AU does not allow us to resolve the degeneracy concerning the physical mechanisms regulating the fragmentation properties. 
    Extreme cases with more than $\sim 11$ fragments appear to be an exception (see Fig. \ref{fig:n_frag_vs_LoverM}), as they are only associated with the \textit{$\mu$100} realizations in our synthetic observations.
    According to these findings, clumps with significantly high levels of fragmentation, likely originate from clumps in which the magnetic field is sub-dominant and vice-versa, in line with the observations by e.g., \cite{2015ApJ...799...74P} and \cite{2020A&A...644A..52A}.
    Given this result, we hypothesize that the SQUALO clumps, hosting from $1$ to $9$ fragments, are more likely to correspond to magnetized clumps. \\
    \indent The comparison between fragment multiplicity in the SQUALO clumps and the parameter space covered by the synthetic observations (see Fig. \ref{fig:parameterspace_1}) allows us to evaluate how well the combinations of initial conditions explored in the simulations (see Table \ref{tab:initial_conditions}), together with the post-processing strategy, reproduce the fragmentation properties observed in actual interferometric data.
    We observe that $7$ out of the $13$ SQUALO clumps are well represented in terms of fragmentation level by the overall distributions. 
    Among the outliers, two clumps, exhibiting $3$ and $4$ fragments, respectively, are very close to the distribution. 
    Four other clumps, exhibiting $2$ fragments or no fragmentation, are poorly represented by the synthetic sample.\\
    \indent Unlike actual observations, our synthetic observations do not reproduce clumps with a very low number of fragments:
    only a few maps are characterized by the presence of two fragments, and no cases with a single fragment.
    Since in our data there is an indication that lower values of $\mu$ may lead to fewer fragments at a resolution of $\sim 7000$ AU, observations with one or two fragments may require stronger magnetic fields than those considered in this study. 
    Alternatively, they could suggest the involvement of additional physical mechanisms to explain the observed fragmentation.
    Following the philosophy of the Rosetta Stone approach, we begin investigating this hypothesis by utilizing the aforementioned collection of control models, which lie outside the parameter grid presented in Table \ref{tab:initial_conditions} (see Appendix \ref{Appendix:ootg_models}). \\
    \indent By looking at the overall coverage of the parameter space, we observe that higher levels of fragmentation are generally associated with more advanced evolutionary stages. 
    This pattern also holds for the SQUALO sample, where high fragmentation is not observed in clumps with low $L/M$. 
    Similar findings, supporting the link between high fragmentation and evolved clumps, have been reported in studies based on $\sim1500$ AU resolution observations, such as those by \cite{2025A&A...696A.151C} and \textcolor{cobalt}{Elia et al. (submitted)}.
    Despite this association, a strict correspondence between the number of fragments and the clump evolutionary stage is not found, either in observations or in synthetic observations. 
    In fact, there are many cases where evolved clumps exhibit low fragmentation levels. 
    This suggests that 
    peculiar interplay of the physical forces and/or the impact of larger-scales dynamics concur to the observed variability in fragmentation levels.\\
    \indent While our synthetic observations predict the formation of up to 14 fragments, higher-resolution observations reveal up to more than 20 fragments (e.g., \citealt{2018A&A...617A.100B}, \citealt{2022A&A...664A..26P}, \citealt{2025A&A...696A.151C}).
    This discrepancy suggests that further fragmentation, occurring at smaller scales that our current observations cannot resolve, is expected.
    As an example, the reference source  HG24 exhibits at $\sim 2000$ AU (\citealt{2025A&A...696A.151C}) one more fragment with respect to the $2$ fragments resolved at $\sim 7000$ AU (\citealt{2023MNRAS.520.2306T}).

    \begin{figure}
        \centering
        \includegraphics[width=0.5\textwidth]{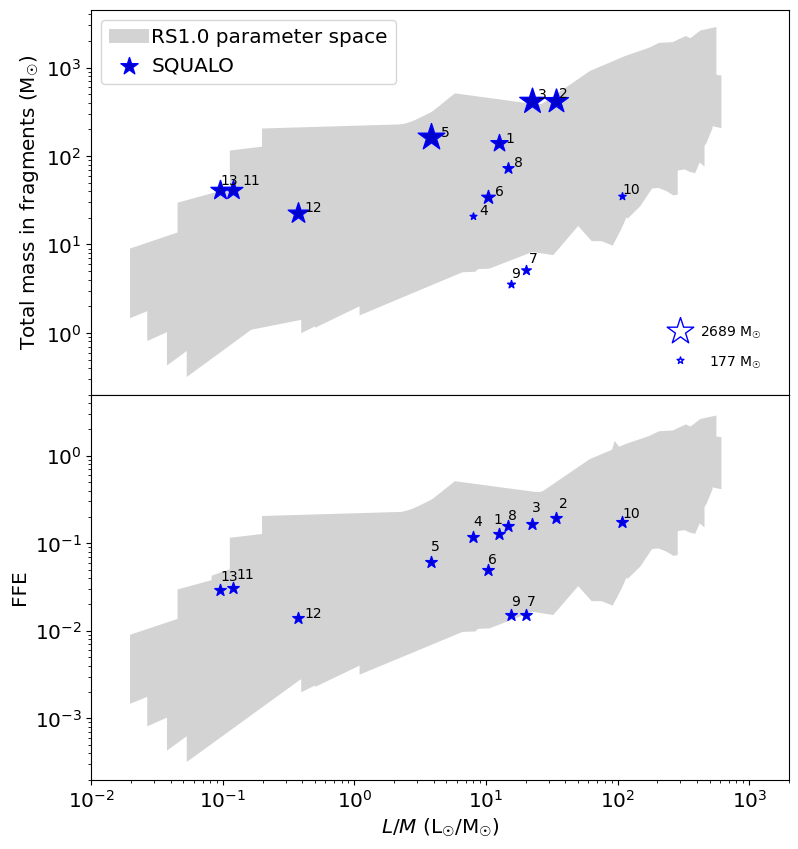}
        
        \caption{Top: Total mass accreted onto the fragments as a function of the clump $L/M$ and comparison with the SQUALO sample. 
        Bottom: FFE as a function of the clump $L/M$ and comparison with the SQUALO sample.
        All the combinations of parameters presented in Table \ref{tab:initial_conditions} are included in the gray area. 
        The stars mark the total mass accreted on fragments for the $13$ clumps of the SQUALO sample, and the relative FFE.  
        In the top panel, the total mass in fragments is size-coded according to the SQUALO clump mass. 
        We also provide a reference of the clump\,mass\,-\,marker\,size relation for the most and least massive SQUALO clumps, of 2689 and 177 M$_\odot$, respectively.}
        \label{fig:parameterspace_2}
    \end{figure}

\subsubsection{Comparison with mass accretion and fragment formation efficiency from the SQUALO sample}

    Building on prior findings, we now compare the total mass in fragments and the FFE of the SQUALO clumps with the parameter space covered by the synthetic observations.
    As observed in Section \ref{Sect:Accretion of mass onto the fragments}, the fragments do not evolve in isolation, but remain closely linked to the parent clump throughout evolution, supporting the clump-fed scenario for star formation (e.g., \citealt{Bonnell1997,Bonnell2001},  \citealt{Klessen2000}, \citealt{2006MNRAS.370..488B},  \citealt{Girichidis2011,Girichidis2012},  
    \citealt{2012A&A...543L...3H},  
    \citealt{2019MNRAS.490.3061V}, 
    \citealt{2020MNRAS.496.3482P},
    \citealt{2023MNRAS.520.2306T}).\\
    \indent In Fig. \ref{fig:parameterspace_2} (top panel), which shows the total mass accreted onto the fragments as a function of the $L/M$ ratio, only a few outliers (and not too far from the overall covered parameter space) were observed among the $13$ SQUALO clumps. 
    The discrepancy likely arises from differences in the initial clump mass between the SQUALO clumps (ranging from 177 to 2689 M$_\odot$) and the synthetic models, which use two fixed initial mass values.   
    In this case, the normalization by the parent clump mass offers a more direct and meaningful way to compare the two datasets.
    Indeed, in Fig. \ref{fig:parameterspace_2} (bottom panel),  
    the FFE values from the synthetic models align well with the entire SQUALO sample, suggesting consistent trends across datasets.
    We acknowledge that the different distances of the SQUALO sources introduce an observational bias in the process of fragment mass estimation (see Eq.\,\ref{eq:mass_f}), though this bias should not be significant (except for sensitivity issues) when normalizing the mass in fragments by the mass of the parent clump (which is computed as a function of the same distance).
    Beside the distance correction, the agreement between datasets implies that the parameter space explored in the RS1.0 suite of simulations successfully reproduces the SQUALO sample in terms of FFE, and further supports the agreement with the clump-fed scenario of star formation already proposed for the SQUALO sources by \cite{2023MNRAS.520.2306T}.
    
\subsubsection{Comparison with relative distances from the SQUALO sample}

\begin{figure}
        \centering
        \includegraphics[width=0.513\textwidth]{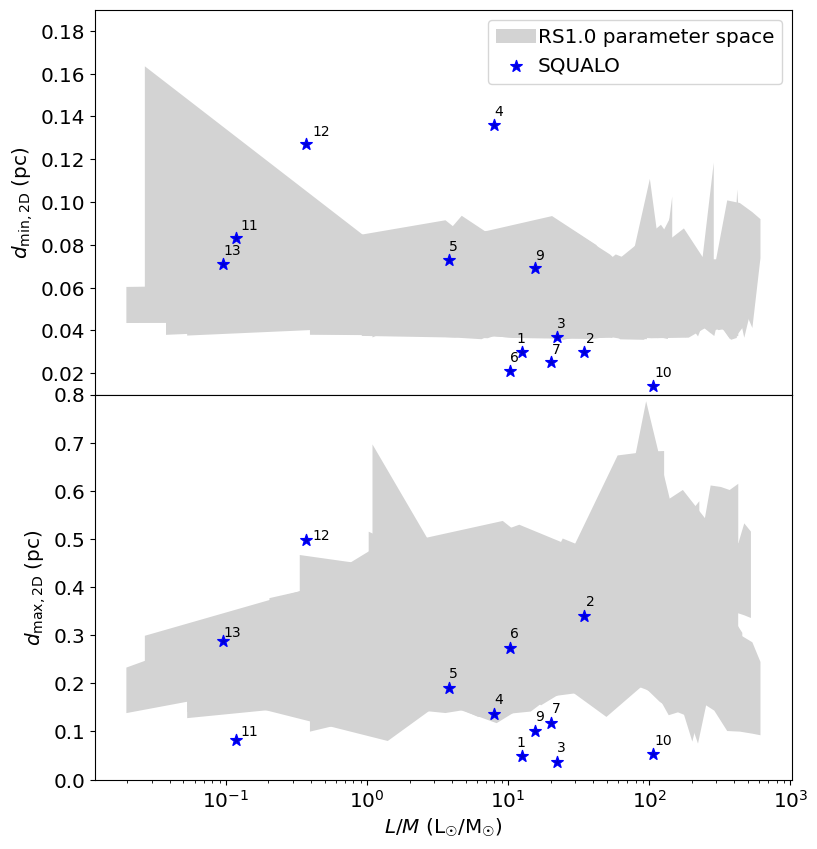}
        \caption{Top: Minimum distance between fragments as a function of the clump $L/M$ and comparison with the SQUALO sample. 
        Bottom: Maximum distance between fragments as a function of the clump $L/M$ and comparison with the SQUALO sample.
        All the combinations of parameters presented in Table \ref{tab:initial_conditions} are included in the gray area. 
        The stars mark the minimum and maximum distances between fragments for the $13$ clumps of the SQUALO sample.}
        \label{fig:parameterspace_3}
    \end{figure}

    When comparing the relative distances between fragments from the SQUALO sample with the parameter space covered by the synthetic observations, we observe some discrepancies across datasets regarding both $d_{\text{min,2D}}$ and $d_{\text{max,2D}}$.
    The associated trends are shown in Fig. \ref{fig:parameterspace_3}.
    In particular, the $d_{\text{min,2D}}$ distribution in the SQUALO sample shows a decreasing trend over time, which contrasts with our findings (see also the comparison between the SQUALO and CORE surveys presented by \citealt{2023MNRAS.520.2306T}).\\
    \indent In our synthetic observations, we do not observe a decrease in $d_{\text{min,2D}}$.
    The large initial spread is due to very high uncertainties (across projections) in the early evolutionary stages of a single realization (see Fig. \ref{fig:Seed1_all_vs_LoverM} for details).
    These uncertainties cover a shaded area that may give the false impression of a general decreasing trend, even though this behavior does not have a match across all the other realizations. 
    The distribution associated with the synthetic observations cannot fully cover all the SQUALO data due to a known resolution issue. 
    Specifically, we are replicating the observations of the farthest source within the SQUALO sample, which corresponds to the lowest linear resolution in that dataset. 
    As a result, in the comparison, we observe SQUALO points falling below our resolution threshold, reflecting the limitations in resolving smaller-scale structures (which we would resolve by mimicking the high-resolution end of the sample). 
    The resolution issue limits our ability to accurately capture the relative distances between fragments, and to compare these values with those obtained from observations at different resolutions.\\
    \indent The $d_{\text{max,2D}}$ distribution should, in principle, not be directly affected by resolution limits. 
    However, the synthetic observations do not accurately reproduce the spread observed in the SQUALO sample, especially for the most compact sources. 
    In this case, $d_{\text{max,2D}}$ is likely not recovered properly due to the absence of large-scale dynamics in our simulations, as we are simulating clumps within isolated boxes. 
    The larger-scale interactions, such as the global collapse of a filament or the interactions with neighboring structures, can significantly influence the morphology of the clumps and the clustering of fragments therein. 
    Without these effects, the fragment distribution in our synthetic clumps seems to be on average less compact than in the SQUALO sample. 
    The presence of feedback mechanisms might as well affect the observed clustering properties.
    We plan to address these limitations in the future RS2.0 suite of models by simulating the collapse of non-isolated clumps incorporating feedback mechanisms.

\section{Summary and conclusions}\label{Sect:Conclusions}

    In this paper, we have presented the general workflow of the Rosetta Stone project: an end-to-end framework that aims to probe the fragmentation process of massive clumps by comparing observations and dedicated numerical simulations using realistic synthetic observations (refer to Paper\,I + Paper\,II + Paper\,III for the complete RS1.0 framework).
    Specifically, we have focused on the comparison of the post-processed simulations produced for this work
    with real ALMA 1.3\,mm observations conducted as part of the SQUALO survey by \cite{2023MNRAS.520.2306T}.

    We have demonstrated that, to effectively address the physics governing star-forming sites at the scales probed by actual observations, it is not only necessary to produce readable synthetic observations, but also to ensure that these mimic the observational features (instrumental effects and observational biases) of the dataset they are compared against. 
    This approach has allowed us to perform a self-consistent and systematic comparison between actual observations and synthetic observations to provide a quantitative characterization of how the initial conditions of clumps and environment affect the observed clump fragmentation properties.
    This comparison was made possible by the estimation of the $L/M$ evolutionary parameter, which was computed for the synthetic clumps in Paper II.

    By the analysis of the fragmentation properties of synthetic SQUALO-like clumps performed at $\sim 7000$ AU resolution, we found the following:
    \begin{itemize}
        \item Fragmentation occurs within the clumps at each evolutionary stage. 
        The fragmentation level observed within individual fields ranges from $2$ to $14$ fragments, with an average of $\sim7$ fragments, and is comparable to the fragment multiplicity recovered in SQUALO (1-9).
        As of now, our suite of simulations does not systematically reproduce cases with $1$ (zero occurrence) or $2$ fragments, which are instead observed in the SQUALO sample.
        \item  At a fixed linear resolution, the number of identified fragments mainly depends on the magnetization of the clump, with high levels of fragmentation preferentially associated with clumps in which the magnetic field is sub-dominant, and vice-versa. 
        This effect is particularly evident in advanced stages of clump collapse corresponding to $L/M\gtrsim20$ L$_\odot$/M$_\odot$.
        In some instances, stochastic processes limit our ability to discern the impact of clump magnetization on the fragments multiplicity.
        \item It is possible that magnetic fields are modulating the fragmentation process also in terms of fragmentation patterns, leading to more tightly bound clump configurations and fragment distributions, i.e. to smaller $d_{\text{max,2D}}$ between fragments, in the evolved magnetized clumps. 
        \item The $d_{\text{min,2D}}$ does not exhibit any trend with evolution. 
        However, this result might be strongly biased by the available linear resolution of $\sim0.036$ pc, which is the minimum observable source separation. 
        \item Among identified fragments, $\sim75\%$ have a match with one or multiple sink particles. This evidence supports a hierarchical and/or multi-layered fragmentation process.
        \item Among the identified fragments that do not have a counterpart in sink particles ($\sim25\%$): some will develop one (or more) sink(s) with the evolution, some are transients, and some others originate from the convolution of the beam with diffuse gas along the line of sight. 
        In the last scenario, identified fragments do not correspond to real compact structures.
        \item Our synthetic observations support a clump-fed star-formation scenario in which fragments are not isolated from the environment. 
        The accretion of mass onto the fragments continues throughout the whole clump evolution.
        Based on the comparison with the SFE, the efficiency of the mass accretion process appears to be scale dependent.
        This represents further evidence of the hierarchical nature of the fragmentation process.
        \item Overall, the parameter space of fragmentation properties explored with the RS1.0 suite of simulations is in good agreement with the SQUALO observations.
        The partial mismatch concerning the number of identified fragments suggests that other physical processes with respect to those explored in our grid may be at play in these particular clumps, and has been the primary motivation for developing a set of control models (see Appendix \ref{Appendix:ootg_models}) outside the grid presented in Table\,\ref{tab:initial_conditions}. 
    \end{itemize}

    As anticipated, the upcoming Rosetta Stone catalogs will feature simulations with a broader parameter space, incorporating more physical effects (such as jets and H{\sc ii} regions) and a more extensive range of scales (spanning from kpc scales to protostellar disks scales). 
    Thanks to its flexibility, the post-processing pipeline will be also tuned accordingly.
    We hope that this end-to-end framework will inspire the community to more systematically compare models and observations by means of synthetic observations to advance our understanding of star formation across all scales.

\begin{acknowledgements}
We thank the anonymous referee for the useful feedback, which has helped to improve the quality of the paper. 
This project was funded by the European Research Council via the ERC Synergy Grant ``ECOGAL'' (project ID 855130). We thank the whole consortium for the stimulating discussions which helped us tremendously through this process.
AT gratefully acknowledges support from a mini-grant funded by INAF. 
This paper makes use of the following ALMA data:
ADS/JAO.ALMA\#2018.1.00443.S. ALMA is a partnership of ESO (representing its member states), NSF (USA) and NINS (Japan), together with NRC (Canada), MOST and ASIAA (Taiwan), and KASI (Republic of Korea), in cooperation with the Republic of Chile.
The Joint ALMA Observatory is operated by ESO, AUI/NRAO, and NAOJ.
We acknowledge PRACE for awarding us access to the JUWELS supercomputer. This work was also granted access to HPC resources of CINES and CCRT under the allocation A0130407023 made by GENCI (Grand Equipement National de Calcul Intensif).
RSK also acknowledges financial support from the German Excellence Strategy via the Heidelberg Cluster ``STRUCTURES'' (EXC 2181 - 390900948) and from the German Ministry for Economic Affairs and Climate Action in project ``MAINN'' (funding ID 50OO2206).  RSK thanks the 2024/25 Class of Radcliffe Fellows for highly interesting and stimulating discussions. 
G.A.F gratefully acknowledges the Deutsche Forschungsgemeinschaft (DFG) for funding through SFB 1601 ``Habitats of massive stars across cosmic time'' (sub-project B1) and support from the University of Cologne and its Global Faculty programme. 
D.W. acknowledges support from the Science and Technology Facilities Council (STFC) [grant number ST/Y004108/1].
\end{acknowledgements}
    
\bibliographystyle{aa_url}
\bibliography{RSIII}
\renewcommand{\arraystretch}{1.3}
\begin{appendix} \label{Sect:Appendix}
\onecolumn
\section{The SQUALO sample}\label{SQUALO}

    In this Appendix we report the table (see Table \ref{tab:SQUALO}) with all the properties of the $13$ clumps from the SQUALO survey (\citealt{2023MNRAS.520.2306T}). 
    Specifically, the clump properties have been taken from the latest release of the Hi-GAL clump catalog (\citealt{2021MNRAS.504.2742E}) and fragment-scale properties have been taken from \cite{2023MNRAS.520.2306T} or derived from the analysis therein. 

    \vspace{0.5cm}

    \begin{table*}[h!]
        \caption{Properties of the $13$ clumps selected in the SQUALO survey. }
        \centering
        \begin{tabular}{lcccc|cccccc}
        \hline
        \addlinespace[2.5pt]
        Designation & $M_\textrm{clump}$ & $L_\textrm{clump}$ & $L_\textrm{clump}$/$M_\textrm{clump}$ & $d$ & $\#$f & $M_{\textrm{tot, fragments}}$ & FFE & $d_{\textrm{min}}$ & $d_{\textrm{max}}$ & r.m.s. \\
        & (M$_\odot$) & (L$_\odot$) & (L$_\odot$/M$_\odot$) & (kpc) & & (M$_\odot$) & & (pc) & (pc) & (mJy/beam) \\
        \addlinespace[2.5pt]
        \hline \hline
        HIGALBM327.3918+0.1996& 1121& 14146  & 12.6 & 5.16 & 9 & 141.9 & 0.13 & 0.03 & 0.05 & 2.68 \\
        HIGALBM327.4022+0.4449& 2157& 73841  & 34.2 & 4.63 & 5 & 423.5 & 0.38 & 0.03 & 0.34 & 8.67  \\
        HIGALBM331.1314-0.2438& 2467& 54845  & 22.2 & 4.95 & 2 & 413.8 & 0.37 & 0.04 & 0.04 & 11.53  \\
        HIGALBM332.6045-0.1674& 177 & 1405   & 7.9  & 3.07 & 2 & 20.7  & 0.02 & 0.14 & 0.14 & 1.15 \\
        HIGALBM338.9260+0.6340& 2689& 10280  & 3.8  & 4.16 & 4 & 163.4 & 0.15 & 0.07 & 0.19 & 3.26  \\
        HIGALBM341.2149-0.2359& 700 & 7210   & 10.3 & 3.45 & 7 & 34.7  & 0.03 & 0.02 & 0.27 & 1.87  \\
        HIGALBM343.5212-0.5172& 347 & 7039   & 20.3 & 3.03 & 5 & 5.2   & 0.01 & 0.03 & 0.12 & 1.07  \\
        HIGALBM343.7560-0.1629& 461 & 6778   & 14.7 & 2.0 & 1 & 73.4  & 0.07 & - & - & 15.18  \\
        HIGALBM344.1032-0.6609& 230 & 3586   & 15.6 & 2.0 & 3 & 3.5& $<$0.01 & 0.07 & 0.10 & 1.64  \\
        HIGALBM344.2210-0.5932 &202  &21645  &107.0 & 2.0 & 7 & 35.5  & 0.03 & 0.01 & 0.05 & 3.81  \\
        HIGALBM24.0116+0.4897\tablefootmark{*} &1334 & 158    & 0.1  & 5.22 & 2 & 40.8  & 0.04 & 0.08 & 0.08 & 1.42  \\
        HIGALBM28.1957-0.0724 &1598 & 594    & 0.4  & 5.36 & 4 & 22.4  & 0.02 & 0.13 & 0.50 & 0.78  \\
        HIGALBM31.9462+0.0759& 1392& 133     & 0.1  & 5.5 & 4 & 40.8  & 0.04 & 0.07 & 0.29 & 0.83  \\
        \addlinespace[2.5pt]
        \hline
        \end{tabular}
        \label{tab:SQUALO}
        \tablefoot{The clump properties have been taken from the latest release of the Hi-GAL clump catalog (\citealt{2021MNRAS.504.2742E}) and fragment-scale properties have been taken from \cite{2023MNRAS.520.2306T} or derived from the analysis therein. 
        Col.1: Source designation ID. Cols. 2–3: Mass $M_\textrm{clump}$ and luminosity $L_\textrm{clump}$ of the parent clumps. Col. 4: $L_\textrm{clump}$/$M_\textrm{clump}$ of the clump, used as evolutionary indicator. 
        Col. 5: Source distances adopted in \cite{2023MNRAS.520.2306T}. 
        Col. 6: Number of fragments $\#$f identified in each clump. 
        Cols. 7-8: Total mass accreted onto the fragments $M_{\textrm{tot, fragments}}$ and fragment formation efficiency FFE obtained by dividing by the mass of the parent clump (normalized to 1). 
        Cols. 9-10: Minimum $d_{\textrm{min}}$ and maximum $d_{\textrm{max}}$ distances between the fragments. 
        Col. 11: Achieved r.m.s..\\\\
        \tablefoottext{*}{HIGALBM24.0116+0.4897 is the reference source for the observational set-up and strategy replicated in this work.}}
    \end{table*}
    
\clearpage
    
\section{Noise consistency analysis}\label{Sect:consistency_checks}

\begin{figure*}[h!]
        \centering
        \includegraphics[width=0.55\textwidth]{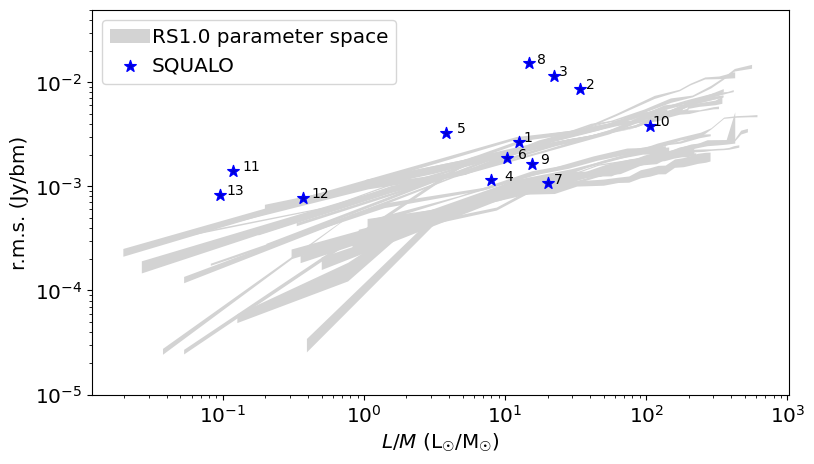}
        \caption{Root mean square level obtained from \textit{Hyper} as a function of the clump $L/M$ and comparison with the SQUALO sample. 
        All the combinations of parameters presented in Table \ref{tab:initial_conditions} are included in the gray area.
        The stars mark the r.m.s. level of the $13$ clumps of the SQUALO sample computed by \cite{2023MNRAS.520.2306T}. 
        }
        \label{fig:parameterspace_4}
    \end{figure*}

    In order to ensure data consistency between the SQUALO reference observations and the synthetic data, we compare the r.m.s. properties obtained for the two datasets (see Fig.
    \ref{fig:parameterspace_4}).
    The estimated r.m.s. for the set of synthetic observations, calculated using the \textit{Hyper} software directly on the cleaned (primary beam non-corrected) maps, ranges from $\sim0.02$ to $\sim15$ mJy/beam. 
    While its values varies significantly across the sample, it is comparable to the r.m.s. range of the cleaned reference observations, which spans from $\sim0.8$ to $\sim15$ mJy/beam. 
    The mean r.m.s. of 2.9 mJy/beam across the sample is consistent with the 1.42\,mJy/beam r.m.s. of the HG24 reference source.

    Notably, the r.m.s. values for the synthetic sample increases with evolutionary stage.
    This effect may be possibly linked to the variation in clump brightness and morphology over time. 
    The integrated signal reflects both the increased flux from compact sources and the ongoing collapse of filamentary structures; indeed, these structures become progressively brighter and less diffuse as the clumps evolve. 
    Despite the $\sim3$-order-of-magnitude span in r.m.s. values, the corresponding dynamic range, computed as the ratio between the peak flux of the brightest source and the r.m.s. of each field,  varies by less than $\sim2$ orders of magnitude.
    The dynamic range estimated across all realizations and all evolutionary steps goes from $\sim2$ to $\sim115$. 
    This relatively narrow range suggests that the increase in peak source brightness and in the background fluctuations are correlated.  
    
\section{Pipelines comparison}\label{Sec:Pipelines comparison}

    In Fig. \ref{fig:model_vs_synt_vs_simalma}, we present a visual comparison that illustrates the impact of different post-processing pipelines on the resulting synthetic images. 
    Specifically, we compare the results obtained using the Rosetta Stone pipeline - which includes a tailored computation of visibilities and the {\ttfamily {tclean}} parameters of the SQUALO project, optimized for clump fragmentation observations - with those from the \texttt{simalma} pipeline implemented in the CASA software.
    This comparison is based on two distinct evolutionary stages of clump collapse, corresponding to $L/M \sim 10$ and 250 L$_\odot$/M$_\odot$, respectively.\\
    \indent
    As mentioned in Section \ref{Sect:Synthetic observations}, if not tuned to reproduce a specific observation, the \texttt{simalma} pipeline operates under default conditions such as ideal instrumental and weather parameters, and continuous integration time.
    The default settings also include general-purpose data reduction and cleaning steps, which are well suited for a wide range of applications.
    However, for our specific science goals, a more customized approach was necessary to fully reproduce the complexity of our reference observations.
    Taking as a reference the outputs from the Rosetta Stone (SQUALO) pipeline, the \texttt{simalma} setup causes a loss of flux from the large-scale emission up to $\sim80\%$ ($\sim40\%$) toward the overall clump structure and up to $\sim50\%$ ($\sim20\%$) toward the individual fragments in the earliest (latest) evolutionary stage. 
    This second estimation is possible only for the cases in which the same fragment is detected in both maps.
    It is likely that a general-purpose algorithm fails to recover most large-scale emission from the \texttt{7\,m} visibilities causing, for example, the presence of dips surrounding the main structure in the most evolved clumps (typically indicative of missing short-spacings). 
    
    Two main factors contribute to the observed flux loss. 
    First, the synthetic acquisition of visibilities in the \texttt{simalma} pipeline employs a single-track observation strategy, leading to a different UV-plane sampling with respect to the coverage achieved with multiple execution blocks in real observations.
    The impact of the different sampling also affects the size of the synthetic beam: in the \texttt{simalma} case the beam is $\sim 1.2$ times the SQUALO one, while the one obtained with the Rosetta Stone pipeline is comparable to our reference.
    Second, the cleaning procedure in \texttt{simalma} lacks tools such as auto-masking and multi-scale cleaning, which are typically used to recover large-scale emission components in the final cleaned image when properly tuned.
    The use of different execution blocks to sample the UV-plane, the incorporation of realistic instrumental setups, the modeling of weather conditions, and the use of joint deconvolution with cleaning and masking techniques implemented by \cite{2023MNRAS.520.2306T}, make the Rosetta Stone pipeline the optimal pipeline for interpreting the reference observations using synthetic observations as a tool.\\
    \indent
    Following the production of realistic synthetic maps, the next step in our methodology involves extracting candidate sources from the images and assessing their photometric properties, which allow us to explore the fragmentation properties of the sample (see Section \ref{Sect:Analysis}). 
    An example of source extraction for the fields in Fig. \ref{fig:model_vs_synt_vs_simalma} is provided to demonstrate how the choice of pipeline also affects source identification. 
    In particular, the different UV-plane coverage and deconvolution strategies significantly affect the quality of the cleaned images, and consequently, the accuracy of source identification.
    This is particularly evident in the early stages of clump collapse. 
    In the \texttt{simalma} cleaned map at $L/M \sim 10$ L$_\odot$/M$_\odot$, the unrealistic UV coverage leads to the production of false positives and negatives (see Fig.\,\ref{fig:model_vs_synt_vs_simalma}), which make source identification and photometry unreliable unless the source extraction algorithm is carefully tuned. 
    The $L/M \sim 10$ L$_\odot$/M$_\odot$ \texttt{simalma} cleaned image also exhibits a candidate fragment (peaks exceeding three times the local r.m.s.) 
    outside the main emitting structure (gray contours, identified in the model). 
    Two other false candidate fragments are identified outside of this map cut.
    With the results from this comparison, we further validate the data reduction and the production of final images with the fine-tuned Rosetta Stone pipeline.

    \begin{figure}[h!]
        
        \begin{adjustbox}{center}
        \includegraphics[width=1.2\textwidth]{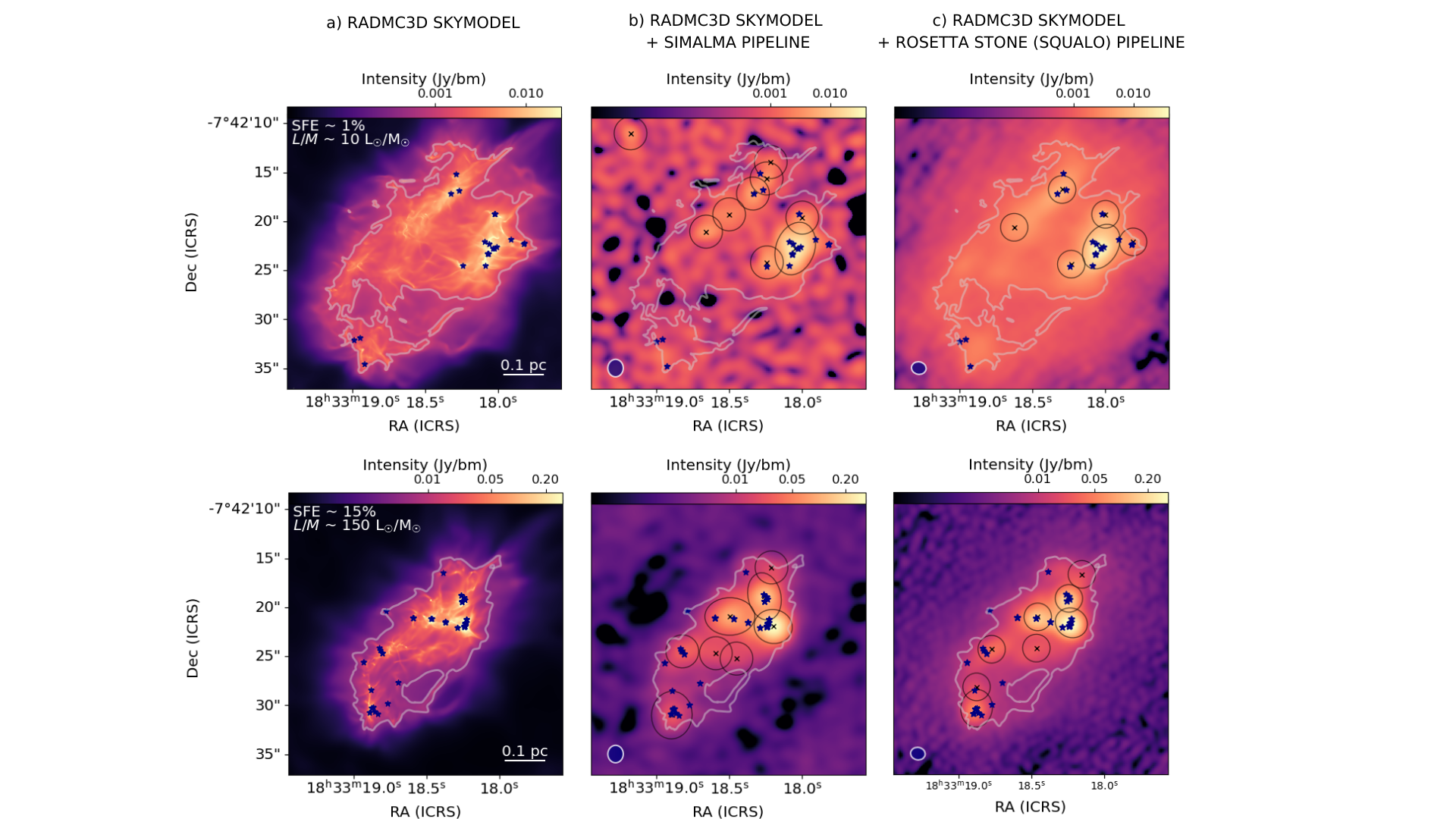}
        \end{adjustbox}
        \caption{Post-processing pipeline comparison chart. Top to bottom: two evolutionary phases of the realization with \textit{M1000\_$\mu$10\_$\mathcal{M}$7\_S2}.
        The snapshots correspond to $\sim 1$ and 15\% SFE, respectively.
        Left to right: a) synthetic surface brightness map after the radiative transfer;
        b) synthetic surface brightness map after the post-processing with the \texttt{simalma} pipeline of the CASA software;
        c) synthetic surface brightness map after the post-processing with the carefully tuned CASA-based Rosetta Stone pipeline, including the SQUALO \texttt{tclean} pipeline.
        The blue stars mark the positions of the sink particles.
        The gray contours mark the main emitting area in the RADMC-3D map and are overlaid on the post-processed maps for comparison.
        The black crosses and ellipses mark the centroids and the contours of the identified fragments.
        Source identification is performed as described in Section \ref{Sect:Source extraction and photometry}.
        In the RADMC-3D map we show the linear scale corresponding to $0.1$ pc, while in the following maps we show the relative synthetic beam footprint.
        }
        \label{fig:model_vs_synt_vs_simalma}
    \end{figure}

    \begin{figure*}[ht]
        \centering
        \renewcommand{\thefigure}{
        D.\arabic{figure}}
        \setcounter{figure}{0} 
        \includegraphics[width=\textwidth]{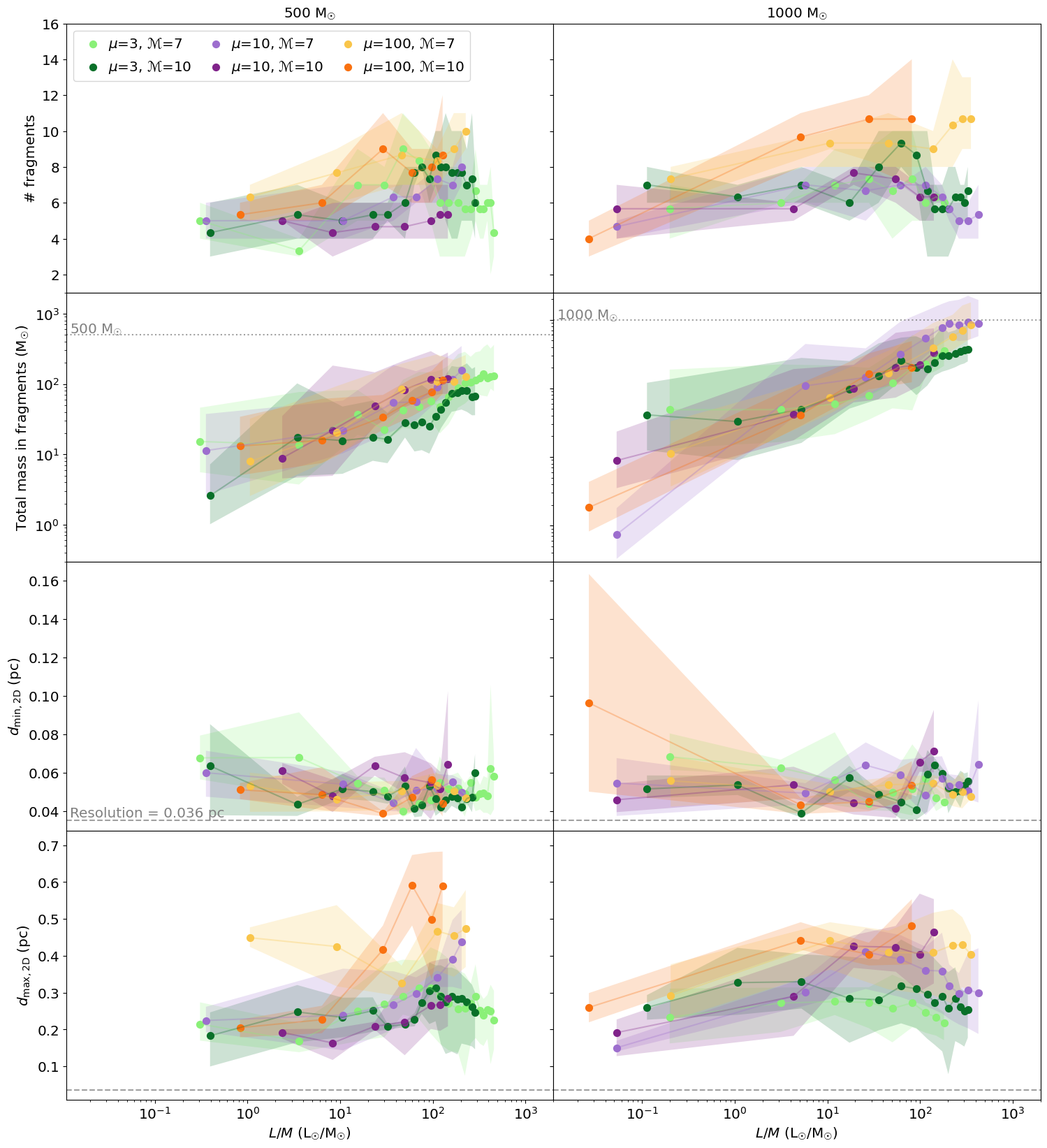}
        \caption{Top to bottom: Number of fragments identified in each clump, total mass accreted onto the fragments, minimum and maximum distances between the fragments as a function of the clump $L/M$. 
        Results from \textit{S1} realizations. Left: Realizations with initial clump mass of 500\,M$_\odot$. 
        Right: Realizations with initial clump mass of 1000\,M$_\odot$.
        Shaded areas illustrate the span of the analyzed quantities across the three projections, indicating the uncertainty in the respective distributions. 
        Only in the case of the total mass in fragments, the shaded areas represent the stochastic variations of fragments temperature due to 1000\,MC runs.
        The color code accounts for the different initial conditions of the realizations, as described in Fig.\,\ref{fig:n_frag_vs_LoverM}.}
        \label{fig:Seed1_all_vs_LoverM}
    \end{figure*}

\section{Fragmentation properties from \textit{S1} realizations}\label{Sect: Seed1 fragmentation properties}
    
    In Fig. \ref{fig:Seed1_all_vs_LoverM}, we provide the distribution of the number of fragments identified in each clump, the total mass in fragments and the relative distance between the fragments as a function of the clump $L/M$ for the \textit{S1} realizations.
    In contrast to \textit{S2} results, the impact of the clump magnetization on the number of fragments is less clear, especially across the \textit{M500} realizations.
    This discrepancy is mainly due to the peculiar stochastic processes at play when the same models are initialized with a different seed.
    In the \textit{S1} case, the stochasticity has larger impact on the fragmentation, and its effects seem to counteract the magnetic regulation.
    However, we cannot rule out the possibility that the combination of the stochastic seed and our fixed orthogonal projections may be allowing us to better address the impact of the physical initial conditions of the simulations for the \textit{S2} realizations. 
    General considerations on the fragmentation properties reported in Section \ref{Sect:Analysis} hold true also for the \textit{S1} realizations.

\section{The necessity of having an end-to-end framework}\label{Appendix:ootg_models}
    
\begin{figure}[h!]
        \centering
        \includegraphics[width=0.55\textwidth]{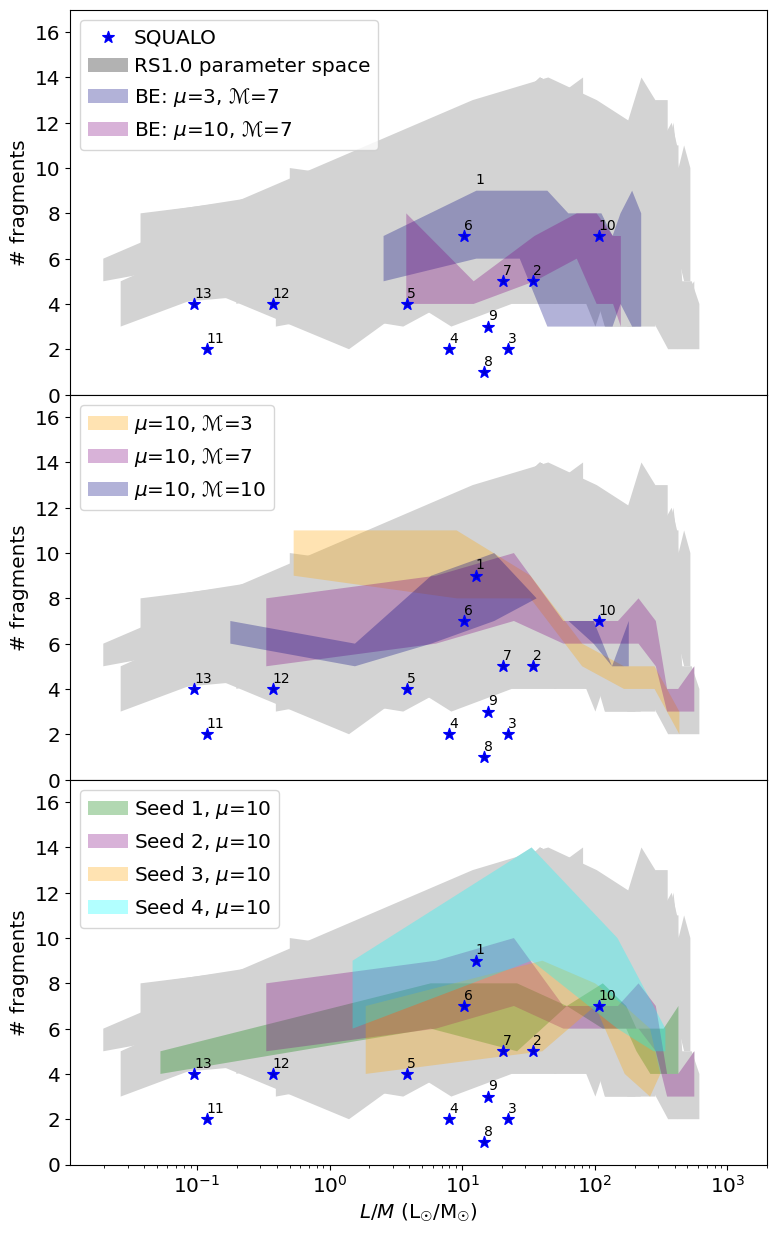}
        \caption{Parameter space exploration of the control models. Number of fragments as function of $L/M$: comparison with the SQUALO sample.
        All the combinations of parameters presented in Table \ref{tab:initial_conditions} are included in the gray area.
        In the top panel, the blue and the purple shaded areas identify two models initialized with a Bonnor-Ebert density profile: \textit{M1000\_$\mu3\_\mathcal{M}7\_S2\_$BE} and \textit{M1000\_$\mu10\_\mathcal{M}7\_S2\_$BE}, respectively.
        In the central panel, the yellow shaded area identifies the \textit{M1000\_$\mu10\_\mathcal{M}3\_$S2} realization produced to test the effect on the fragmentation of a lower value of turbulence.
        It is compared with two models from the RS1.0 suite, i.e., \textit{M1000\_$\mu10\_\mathcal{M}7\_$S2} and \textit{M1000\_$\mu10\_\mathcal{M}10$\_S2}, in purple and blue, respectively. 
        In the bottom panel, the yellow and the cyan shaded areas identify the \textit{M1000\_$\mu10\_\mathcal{M}7$\_S3} and \textit{M1000\_$\mu10\_\mathcal{M}7$\_S4} realizations, respectively, produced to test the effect of of two additional seeds on the fragmentation. 
        They are compared with two models from the RS1.0 suite, i.e., \textit{M1000\_$\mu10\_\mathcal{M}7$\_S1} and \textit{M1000\_$\mu10\_\mathcal{M}7$\_S2}, in green and purple, respectively.}
        \label{fig:parameterspace_5}
    \end{figure}

    Following the philosophy of the Rosetta Stone project, we investigate how to recover clumps with low levels of fragmentation by utilizing a collection of control models that lie outside the parameter grid (see Table \ref{tab:initial_conditions}) used to derive the results presented in the main body of this paper. 
    These different models, shown in Fig. \ref{fig:parameterspace_5}, include two realizations with an initial Bonnor-Ebert (BE) profile, a realization with $\mathcal{M}=3$,
    and two realizations in which we explore the impact of different initial seeds, \textit{S3} and \textit{S4}, respectively. 
    None of these control models allows us to systematically reproduce the cases of low or no fragmentation.
    Only in the case of the Bonnor-Ebert sphere with initial $\mu=3$ (top panel), and in the case with $\mathcal{M}=3$ (central panel), we cover two small portions of the parameter space which are not populated by the RS1.0 models.
    
\begin{figure}[h!]
        \centering
        \includegraphics[width=0.55\textwidth]{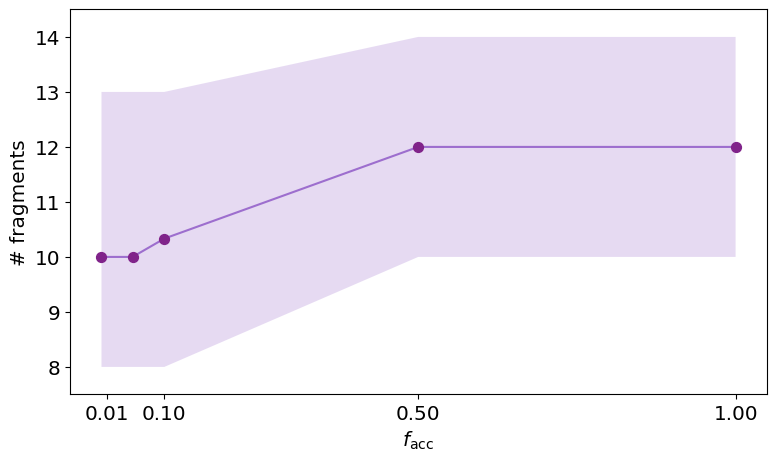}
        \caption{Number of fragments identified in each clump as a function of $f_{\textrm{acc}}$. 
        Shaded areas illustrate the range in number of fragments recovered across the three projections, indicating the uncertainty in the distribution.}
        \label{fig:nfrag_vs_facc}
    \end{figure}

    The case of the comparison of the different seeds is of particular interest (see Fig. \ref{fig:parameterspace_5}, bottom panel). 
    Variations in the initial seed, which influence the morphology of the evolving clumps, typically induce fluctuations of similar magnitude to those associated with projection effects (i.e., variations of a few fragments). 
    However, in some instances the stochastic processes associated with the choice of the initial seed can completely dominate the fragmentation (as in e.g., Fig. \ref{fig:parameterspace_5}, realization \textit{M1000\_$\mu10\_\mathcal{M}7$\_S4}).
    Also in less extreme examples, their impact might hinder the chance to disentangle the role of the other physical initial conditions, as assessed in the case of \textit{M500\_S1} realizations (see Appendix \ref{Sect: Seed1 fragmentation properties}).
    \clearpage
    Beside changing the physical initial conditions, we test the potential impact on the number of observed fragments due to a variation of the numerical prescriptions for the luminosity accretion factor ($f_{\textrm{acc}}$).
    The efficiency of the accretion affects both the thermal feedback from accretion processes and the total luminosity of each sink.
    The test is conducted at an intermediate evolutionary step of the \textit{M1000\_$\mu$100\_$\mathcal{M}$7\_1} realization.
    In our suite of simulations RS1.0, we compute the accretion luminosity using a fixed factor of 0.1. 
    To further investigate the impact of this parameter on the number of detected fragments, we explore four additional values of $f_{\textrm{acc}}$: 0.01, 0.05, 0.5, and 1, respectively. 
    These values represent a broad range of possible luminosity accretion conditions. 
    As a first test, we change $f_{\textrm{acc}}$ in the radiative transfer calculations, using the POLARIS\footnote{POLARIS is not the main software used in the RS1.0 pipeline, but it has been used to carry out this specific test. 
    Before running the test, we have ensured that outputs from the two codes are comparable: POLARIS recovers about 96\% of the total RADMC-3D emission from the box at 1.3 mm. 
    We therefore expect the following results to be representative of the same analysis conducted by using RADMC-3D.} code \citep{Reissl+2016, Reissl+2018} to calculate the temperature and the resulting intensity maps. 
    As the variation of the $f_{\textrm{acc}}$ parameter is introduced only in the radiative transfer phase rather than in the underlying RMHD simulation, it does not directly change the fragmentation of the clumps and the number of sinks. 
    Rather, it modifies our ability to detect the fragments at a resolution of $\sim7000$ AU that may correspond to those sinks.\\
    \indent As shown in Fig. \ref{fig:nfrag_vs_facc}, we observe a slight change in the number of fragments as the $f_{\textrm{acc}}$ parameter is varied. 
    Specifically, higher values of $f_{\textrm{acc}}$ (i.e., 0.5 and $1$) lead to slightly more fragments being detected compared to the lower and intermediate values. This suggests that increased luminosity from accretion luminosity contributes to the identification of $\sim 7000$ AU fragments. 
    We also find that lowering $f_{\textrm{acc}}$ primarily affects the total luminosity of the low-mass sinks, which derive most of their luminosity from accretion luminosity (in contrast to high-mass sinks which are dominated by internal luminosity). 
    As a consequence, in the CASA post-processed images we still detect fragments associated with high-mass stars, while it becomes more challenging to recover the population of the faintest sources linked to lower-mass sinks. 
    Despite the observed variation in the number of fragments, the accretion luminosity factor does not appear to be the primary factor affecting the recovered fragments multiplicity in this context.\\
    \indent Although the low fragmentation cases are not reproduced within the current simulated parameter space, the outputs from both the control models and the RS1.0 grid help refine the initial conditions and the numerical assumptions for our simulations, guiding future investigations.

\end{appendix}

\end{document}